\documentclass[twocolumn,showpacs,amsmath,amssymb,floatfix,superscriptaddress,prd]{revtex4-1}

\usepackage{graphicx}
\usepackage{dcolumn}
\usepackage{bm}
\usepackage{color}
\usepackage{ulem}
\usepackage{multirow}
\usepackage{hyperref}

\def\be{\begin{equation}}
\def\ee{\end{equation}}
\def\beqn{\begin{eqnarray}}
\def\eeqn{\end{eqnarray}}
\def\nn{\nonumber}

\definecolor{darkgreen}{RGB}{0,120,0}

\newcommand{\ground}{Planck+G}

\newcommand\lsim{\mathrel{\rlap{\lower4pt\hbox{\hskip1pt$\sim$}}
        \raise1pt\hbox{$<$}}}
\newcommand\gsim{\mathrel{\rlap{\lower4pt\hbox{\hskip1pt$\sim$}}
        \raise1pt\hbox{$>$}}}

\def\n{{\bf \widehat n}}
\def\ellmax{\ell_{\rm max}}
\def\fsky{f_{\rm sky}}
\def\smallsum{\mathop{\textstyle\sum}\limits}

\begin{document}

\title{Non-Gaussian structure of the lensed CMB power spectra covariance matrix}

\author{Aur{\'e}lien Benoit-L{\'e}vy}
\email{benoitl@iap.fr}
\affiliation{%
 UPMC-CNRS, UMR7095, Institut d'Astrophysique de Paris, 75014 Paris, France
}%

\author{Kendrick M. Smith}
\affiliation{Department of Astrophysical Sciences, Princeton University, Princeton, New Jersey 08544-1001, USA}
\author{Wayne Hu}
\affiliation{Department of Astronomy \& Astrophysics, Kavli Institute for Cosmological Physics,  University of Chicago, Chicago, Illinois 60637, USA
}%

\date{\today}

\begin{abstract}

Gravitational lensing of the Cosmic Microwave Background (CMB) encodes cosmological information in the observed anisotropies of temperature and polarization. Accurate extraction of this additional information requires precise modeling of the covariance matrix of the power spectra of observed CMB fields. We introduce a new analytical model to describe the non-Gaussian structure of this covariance matrix and display the importance of second-order terms that were previously neglected.  When compared with direct numerical simulations our model captures parameter errors to better than a few percent for cases where the non-Gaussianity causes an order unity degradation in errors. We also provide a detailed comparison between the information content of lensed CMB power spectra and ideal reconstruction of the lensing potential. We illustrate the impact of the non-Gaussian terms in the power spectrum covariance by providing Fisher errors on the sum of the masses of the neutrinos, the dark energy equation of state, and the curvature of the Universe. 

\end{abstract}

\maketitle

\section{Introduction}

Gravitational potentials of large-scale structure generate a deflection of the trajectories of the Cosmic Microwave Background (CMB) photons, an effect known as CMB lensing \cite{Blanchard87, Bernardeau98, Zaldarriaga99} (see  \cite{Lewis06} for a review).
After its initial detection in cross-correlation with large-scale structure \cite{Smith07,Hirata08}, 
CMB lensing has now been detected with high significance in high-resolution observations from the Atacama Cosmology Telescope \cite{Das2011,Sherwin11} and the South Pole Telescope \cite{vanEngelen12}.

CMB lensing generates a characteristic statistical signature that makes the CMB sensitive to cosmological parameters which directly
influence the growth of cosmic structure.
This breaks the angular diameter degeneracy in the unlensed CMB and improves constraints on parameters such as 
neutrino masses, the dark energy equation of state, and the curvature of the Universe \cite{Metcalf:1997ad, Stompor:1998zj, Hu02_synergy, Kaplinghat03, Acquaviva06}.

Mathematically, CMB lensing is described as follows.
We introduce a vector field ${\bf d}(\n)$ (the deflection field) such that the lensed temperature $T(\n)$ and
unlensed temperature $\tilde T(\n)$ are related by
\be
T(\n) = \tilde T(\n + {\bf d}(\n))
\ee
and analogously for the Stokes parameters $Q(\n)$, $U(\n)$ which describe linear CMB polarization.
To lowest order in perturbation theory, the deflection field ${\bf d}(\n)$ is the gradient of a scalar lensing potential (i.e.~${\bf d}(\n) = \nabla\phi(\n)$)
which can be written as a line-of-sight integral:
\be
\phi(\n) = -2 \int d\eta \frac{\chi(\eta-\eta_{\rm{rec}})}{\chi(\eta_{\rm{rec}})\chi(\eta)}\Psi(\chi \n, \eta),
\ee
where $\Psi$ is the Newtonian potential, $\eta$ is conformal time, $\eta_{\rm{rec}}$ is the epoch of last scattering, and  $\chi$ is the angular diameter distance in comoving coordinates.

CMB lensing modifies the Gaussian statistics of the unlensed CMB by generating a correlation between the primary field and its gradient \cite{Hu00}.
It also modifies the shape of the temperature and $E$-mode polarization power spectra, and generates a nonzero $B$-mode power spectrum.
This leads to two different statistical techniques for extracting cosmological information from CMB lensing.
First, we can simply make precise measurements of CMB power spectra (especially the $B$-mode power spectrum), which will include
lensing contributions.
Second, we can reconstruct the lensing potential $\phi$ using correlations between the primary and its gradient, providing a new cosmological observable \cite{Benabed:2000jt,Guzik:2000ju,Hu:2001tn,Hu02,Hirata03}.

Accurate analysis of CMB anisotropies requires correct modeling of the covariance matrix of the lensed power spectra.
The lensed CMB is not a Gaussian field, and so its power spectrum covariance is nontrivial; in particular the off-diagonal
correlations are important.
Calculations of the non-Gaussian covariance of the lensed power spectra have been performed in both flat sky \cite{Smith06, SmithS06} and full sky \cite{Li07} cases, but these calculations make the approximation that some high-order terms in the lensing potential are negligible.

In the advent of low-noise and high-resolution CMB experiments that will be able to probe polarization of the CMB at the arcminute scale (SPTPol, ACTPol \cite{ACTPol10}, POLARBEAR \cite{Polarbear11}), it becomes necessary to assess the validity of the current approximations for the non-Gaussian power spectrum covariance, and study the impact on cosmological parameter estimation. The purpose of this paper is therefore to investigate in detail the impact of the non-Gaussianities induced by CMB lensing and quantify the information contained in the power spectra. We introduce a new semi-analytical approach to compute the power spectrum covariance matrix and validate our model by Monte Carlo simulations. 

In Sec. \ref{sec:pscov}, we describe the simulations we performed to estimate the power spectrum covariance matrix and present our semi-analytical model. In Sec. \ref{sec:lens_info}, we introduce a model independent way of characterizing the relative
information content of lensed power spectra and  idealized direct reconstruction.
Finally, in Sec. \ref{sec:cons}, we apply this characterization to specific cosmological
model parameters.

Throughout the paper, we use a fiducial flat $\Lambda$CDM model with the following parameters:
\beqn
&& \{ \Omega_ch^2, \Omega_bh^2, h, \tau, n_s, 10^9 A_s, \smallsum m_\nu \} \\
&& \hspace{0.3cm} = \{ 0.1096, 0.0226, 0.693, 0.089, 0.964, 2.419, 0.58\mbox{ eV} \} .\nn
\eeqn
We chose a large fiducial value of $m_\nu$ so as to be testable with CMB lensing
in the near future.

\section{Non-Gaussian Power Spectrum Covariance}
\label{sec:pscov}

To characterize the information content in CMB power spectra we first require an accurate
characterization of the covariance matrix between the band power estimates of the
temperature and polarization fields.   Previous analytic characterizations have been based
on a perturbative expansion of the effect of lensing \cite{Smith06,Li07} which breaks down
in the damping tail.   In this section, we first conduct a suite of simulations to characterize
the covariance and then develop analytic tools which characterize its main features.

\subsection{Simulations}
\label{ssec:simulations}

We simulate the lensed CMB on the full sky using the following procedure.
We first make a realization of the unlensed CMB fields  $\tilde T$, $\tilde E$ 
and lensing potential $\phi$ in harmonic space,
treating these fields as Gaussian and computing power spectra using the publicly available Boltzmann code CAMB \cite{Lewis:1999bs}
to $\ell_{\rm max}=5000$.
Using a fast spherical harmonic transform, we compute the unlensed temperature
and polarization in pixel space, using an equicylindrical pixelization with
$(N_\theta,N_\phi)=(16384,32768)$ equally spaced points in $(\theta,\phi)$.
We then evaluate the lensed temperature and polarization fields
\be
X(\n) = \tilde X(\n + {\bf d}(\n)),
\ee
where $X\in\{T,Q,U\}$ and tildes distinguish unlensed from lensed fields throughout, at each point of an $N_{\rm side}=4096$ Healpix pixelization \cite{Gorski2005},
using cubic interpolation on the equicylindrical map to evaluate the right-hand side,
and parallel translation of the spin-2 field $Q\pm iU$ to transport polarization at the
point $(\n+{\bf d}(\n))$ to the point $\n$.
(We use an equicylindrical pixelization for the unlensed fields, rather than an irregular
pixelization such as Healpix, so that interpolation is straightforward to implement.)
Taking another spherical transform to obtain
lensed $T$, $E$, and $B$ maps in harmonic space, we compute lensed power spectra $\hat C_\ell^{XY}$
for $XY\in \{TT, TE, EE, BB\}$.
As a memory optimization, we avoid storing full-sky maps by ``striping'' the sky into
16 latitude bands, and calculate the contribution to $a_{\ell m}^{T}$,
$a_{\ell m}^{E}$, $a_{\ell m}^{B}$ from each band before
moving onto the next.
This allows each simulation to fit onto a single core with $\sim$2 GB memory.
The above procedure is similar algorithmically to the publicly available code LensPix~\cite{Lewis2005},
although the two codes differ in minor details of implementation.

We first 
compare the mean over the $N=32768$ realizations
\begin{equation}
\bar C_\ell^{XY} = {1\over N}\sum_{\alpha=1}^N \hat C_{\ell,\alpha}^{XY},
\end{equation}
where $XY \in TT, TE, EE, BB$ to 
the predicted lensed power spectra computed by CAMB.
With the  resolution parameters given above, the lensed CMB power spectra of the simulations agree with CAMB's calculation of the
lensed $C_\ell^{XY}$'s to better than 0.1\% for all spectra at $\ell_{\rm max}=3000$.

We then estimate the covariance matrix between two different power spectra
$XY$ and $ZW$ as
\be
{\rm Cov}^{XY, ZW}_{\ell_1 \ell_2 }= \frac{1}{N}\sum_{\alpha=1}^N \hat C_{\ell_1,\alpha}^{XY} \hat C_{\ell_2,\alpha}^{ZW}-
\bar C_{\ell_1}^{XY} \bar  C_{\ell_2}^{ZW}.
\ee

Even in these noise-free simulations, the Gaussian random variance from the unlensed CMB makes
the estimate of the covariance between individual multipoles noisy.  We therefore
further bin the power spectrum estimators into band powers
\begin{equation}
D^{XY}_i = \sum_\ell B^\ell_i C^{XY}_\ell ,
\label{eqn:CMBband}
\end{equation}
where $B^\ell_i$ is a top hat function
\be
B^\ell_i= \left\{
\begin{array}{ccc}
\frac{1}{\ell_{i+1}- \ell_{i}}&,& \ell_{i} \le \ell<\ell_{i+1}\\
0 &,& \mbox{otherwise}.
\end{array}\right.
\label{eqn:bandwindow}
\ee

The band width is chosen to be sufficiently small so as to resolve the acoustic
features in the spectrum.  In practice we take every multipole to $\ell=25$ followed by 
uniform bands of $\ell_{i+1}-\ell_i=15$. We choose not to bin the 25 first multipoles as the derivatives of the power spectrum with respect to cosmological parameters exhibit strong variation at low multipoles, and averaging these variations to one single bin at low-$\ell$ would give erroneous final results.
The covariance matrix between these band estimators then becomes
\begin{equation}
{\rm Cov}^{XY, ZW}_{ i j } = \sum_{\ell_1,\ell_2} B^{\ell_1}_i {\rm Cov}^{XY, ZW}_{\ell_1\ell_2 }
B^{\ell_2}_j .
\end{equation}

\begin{figure*}[htb]
\vspace{-0.4cm}
\includegraphics[width=1.7\columnwidth]{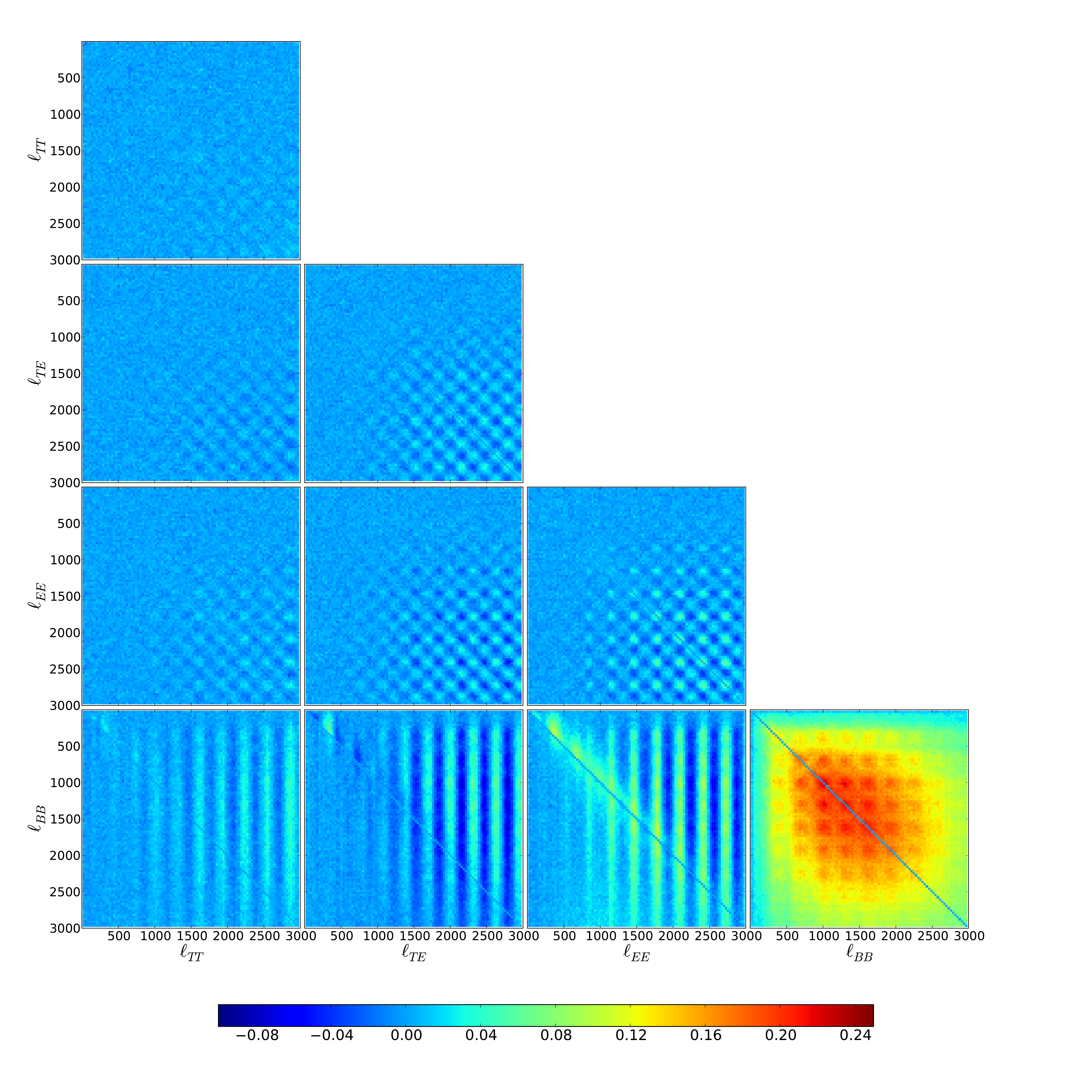}
\caption{\label{Corr_MC} Monte Carlo covariance of the lensed CMB bandpowers computed from 32768 lensed CMB simulations in bands of $\Delta \ell = 30$. From left to right and top to bottom: $TT$, $TE$, $EE$, and $BB$. For visualization purposes we plot the correlation coefficient $R$ defined in Eq.~(\ref{eqn:corr}). The diagonal (of order unity) has been subtracted to enhance contrast. }
\end{figure*}

\begin{figure*}[htb]
\vspace{-0.4cm}
\includegraphics[width=1.7\columnwidth]{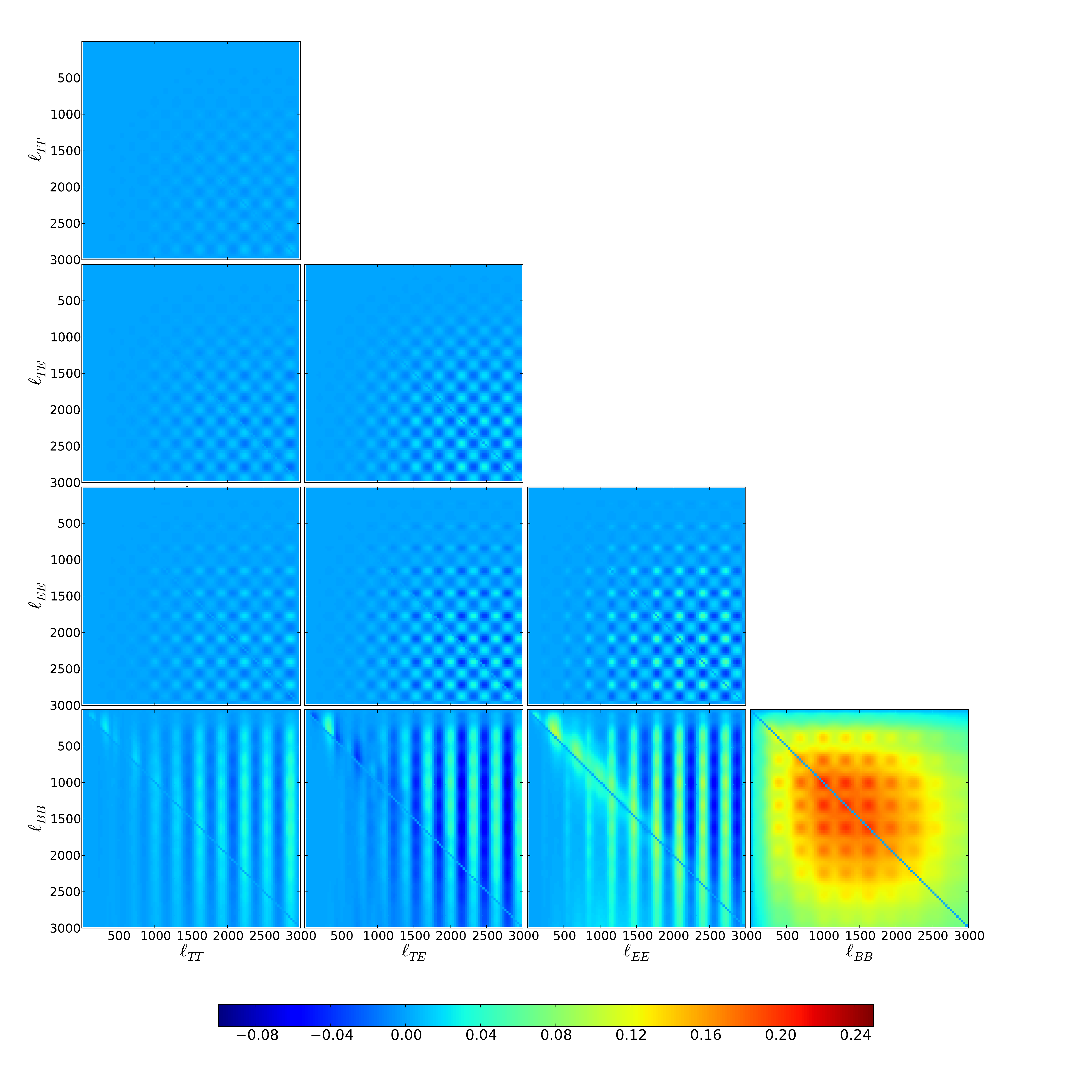}
\caption{\label{Corr_ana2} Covariance of the lensed CMB bandpowers in bands of $\Delta \ell = 30$, computed using the analytic model
from \S\ref{ana_approx}.
 From left to right and top to bottom: $TT$, $TE$, $EE$, and $BB$. For visualization purposes we plot the correlation coefficient $R$ defined in Eq.~(\ref{eqn:corr}). Detailed comparison with the Monte Carlo covariance from Fig.~\ref{Corr_MC} shows that the agreement is excellent.}
\end{figure*}

\begin{figure}[htbp]
\begin{center}
\includegraphics[width=1.00\columnwidth]{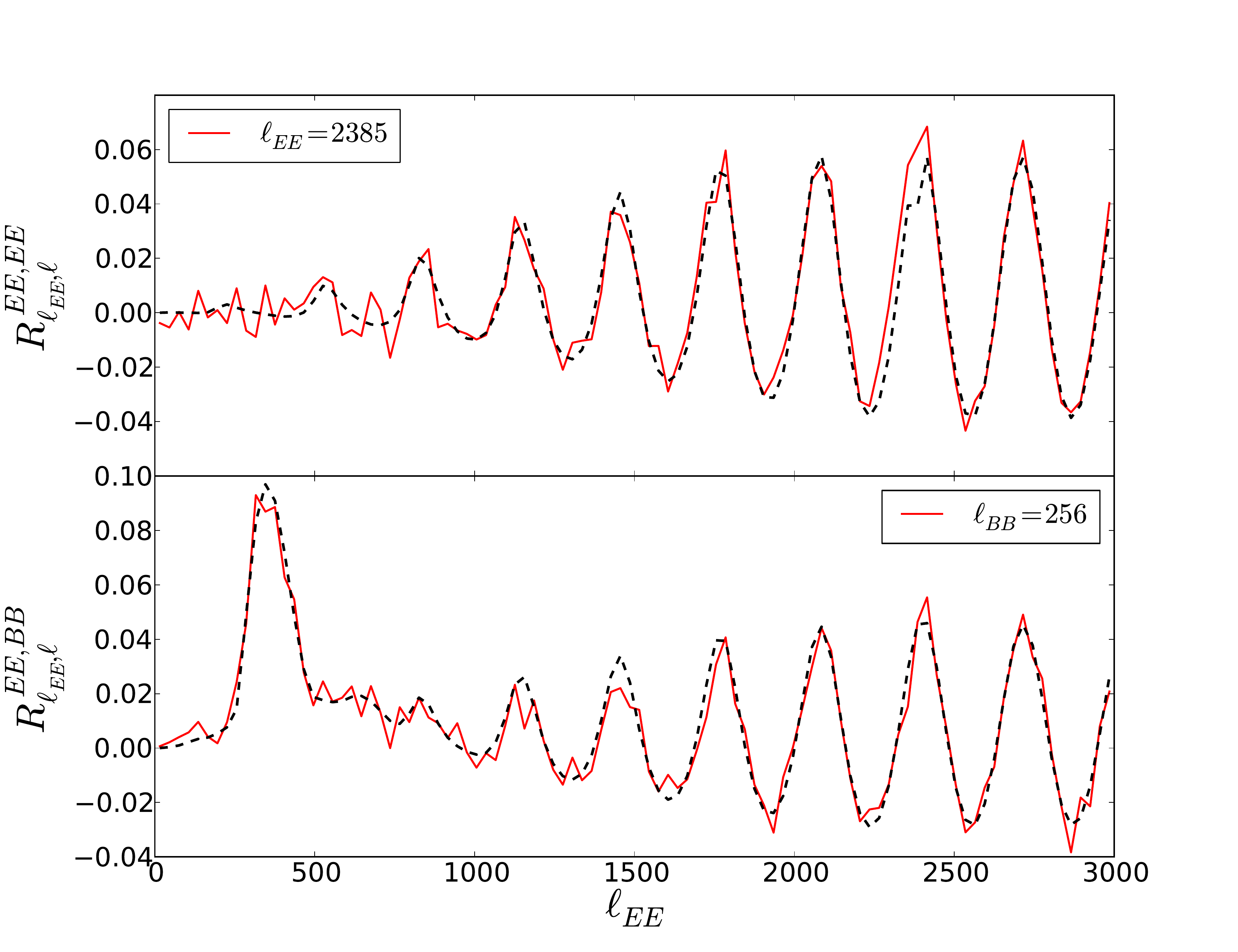}
\end{center}
\caption{ \label{corr1} Rows of the correlation matrix $R$, defined in Eq.~(\ref{eqn:corr}), between $EE$-$EE$ band powers (top), and $EE$-$BB$ (bottom), computed using either Monte Carlo simulations (solid lines) or our analytic model (dashed lines).
Binning scheme follows Fig.~\ref{Corr_MC}  and the autocorrelation is omitted.}
\end{figure}

\begin{figure}[htbp]
\begin{center}
\includegraphics[width=1.00\columnwidth]{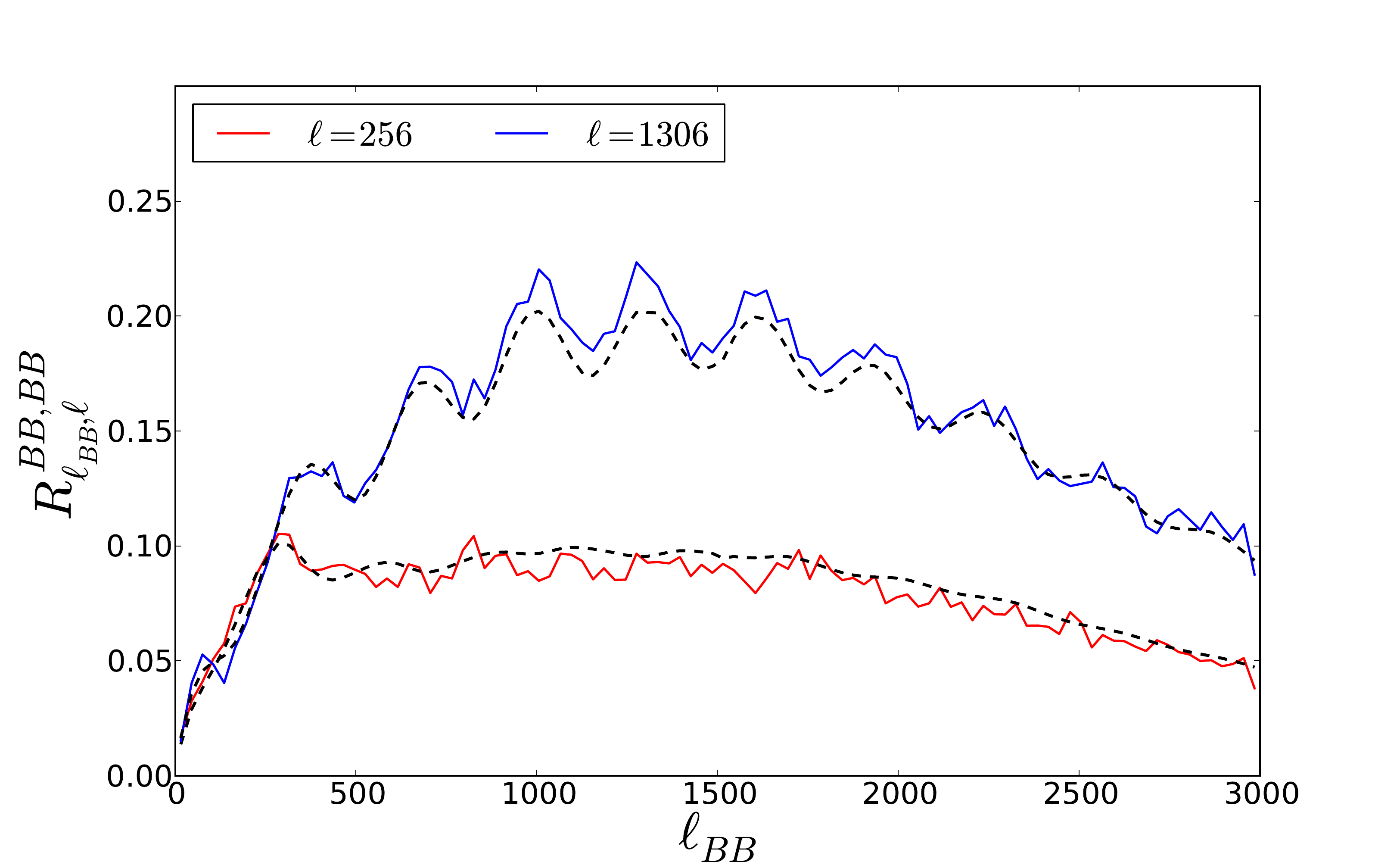}
\end{center}
\caption{ \label{corr} Rows of the correlation matrix $R$, as in Fig.~\ref{corr1}, between $BB$-$BB$ bandpowers computed using either Monte Carlo simulations (solid lines) or our analytic model (dashed lines). Binning scheme follows Fig.~\ref{Corr_MC}  and the autocorrelation is omitted.  The upper curves are for $\ell= 1306$.}
\end{figure}

The Monte Carlo bandpower covariance is shown in Fig.~\ref{Corr_MC}.
For visualization purposes, it
is convenient to scale out the diagonal contributions by defining the correlation matrix
\be
R^{XY, ZW}_{ij}=  \frac{{\rm Cov}^{XY, ZW}_{ij}} {\sqrt{{\rm Cov}^{XY,XY}_{ii} {\rm Cov}^{ZW,ZW}_{jj}}}.
\label{eqn:corr}
\ee
For display purposes, we also use a flat binning scheme by dividing the range of multipoles [2-3000] in 100 bins in Figs.~\ref{Corr_MC}--\ref{corr}.
As expected from previous studies, the covariance of the $B$-modes is highly non-Gaussian \cite{Smith:2004up,Smith06,Li07}.  
Interestingly, the $EE,BB$ and $EE,EE$ correlations in Fig.~\ref{Corr_MC}
are substantially larger than expected from the lowest-order analytic calculations in \cite{Smith06,Li07}, 
and all but $TT,TT$ show clear evidence for correlated structure on the
 acoustic scale that is again not expected.  Although Ref.~\cite{Smith06} also conducted simulation tests, their bands were much wider than the acoustic scale such that  these structures were hidden.

\subsection{\label{ana_approx}Analytic Approximation}

In order to develop a new analytic approximation to the covariance matrix, it is useful to first 
examine the $BB,BB$ correlation for which the existing models work well.  
The dominant terms in the analytic $BB,BB$ correlation expression can be compactly written
as (cf.~\cite{Smith:2004up} Eq.~17) 
\begin{eqnarray}
{\rm Cov}^{BB,BB}_{\ell_1\ell_2} & =& 
 {2 \over 2\ell_1+1} \left(C_{\ell_1}^{BB}\right)^2 \delta_{\ell_1,\ell_2} \nn
\\
&& + \sum_{\ell}\Bigg(
{\partial C_{\ell_1}^{BB}  \over \partial  C_{\ell}^{\tilde E\tilde E} } {\rm Cov}_{\ell\ell}^{\tilde E\tilde E,\tilde E\tilde E}
{\partial C_{\ell_2}^{BB} \over \partial  C_{\ell}^{\tilde E\tilde E} } 
\Bigg)  \nonumber\\ &&
 + \sum_{\ell}\Bigg(  {\partial C_{\ell_1}^{BB}\over \partial  C_{\ell}^{\phi\phi} } {\rm Cov}_{\ell\ell}^{\phi\phi,\phi\phi} {\partial C_{\ell_2}^{BB} \over \partial  C_{\ell}^{\phi\phi} } \Bigg) ,
\label{eqn:BBBB}
\end{eqnarray}
where unlensed CMB power spectra are denoted with tildes and $C_\ell^{\phi\phi}$ is the lensing potential power spectrum.   Assuming that these fields are Gaussian,
we can use the general prescription for Gaussian random fields $G$
\begin{equation}
{\rm Cov}^{G_a G_b,G_c G_d}_{\ell \ell'}  = 
 {\delta_{\ell,\ell'} \over 2\ell+1} [ C_{\ell}^{G_a G_c}C_{\ell}^{G_b G_d} + C_{\ell}^{G_aG_d} C_{\ell}^{G_b G_c}
] 
\end{equation}
for the unlensed CMB and $\phi$ fields.

To calculate power spectrum derivatives such as the ones appearing in Eq.~(\ref{eqn:BBBB}), 
we take finite differences between lensed CMB power spectra computed using CAMB, rather than 
using a perturbative expansion in deflection angles.
Since CAMB's algorithm for computing lensed CMB power spectra includes terms of high order
in deflection angles \cite{Challinor:2005jy}, this approach to computing derivatives
also includes high order terms, and in particular does not break down at high $\ell$.
Some implementational details of the derivative calculation are presented in Appendix~\ref{app:derivatives}.
The model of Eq.~(\ref{eqn:BBBB}) for the correlation matrix is shown in Fig.~\ref{Corr_ana2}.

\begin{figure}[htb]
\vspace{-0.4cm}
\begin{center}
\includegraphics[width=1\columnwidth]{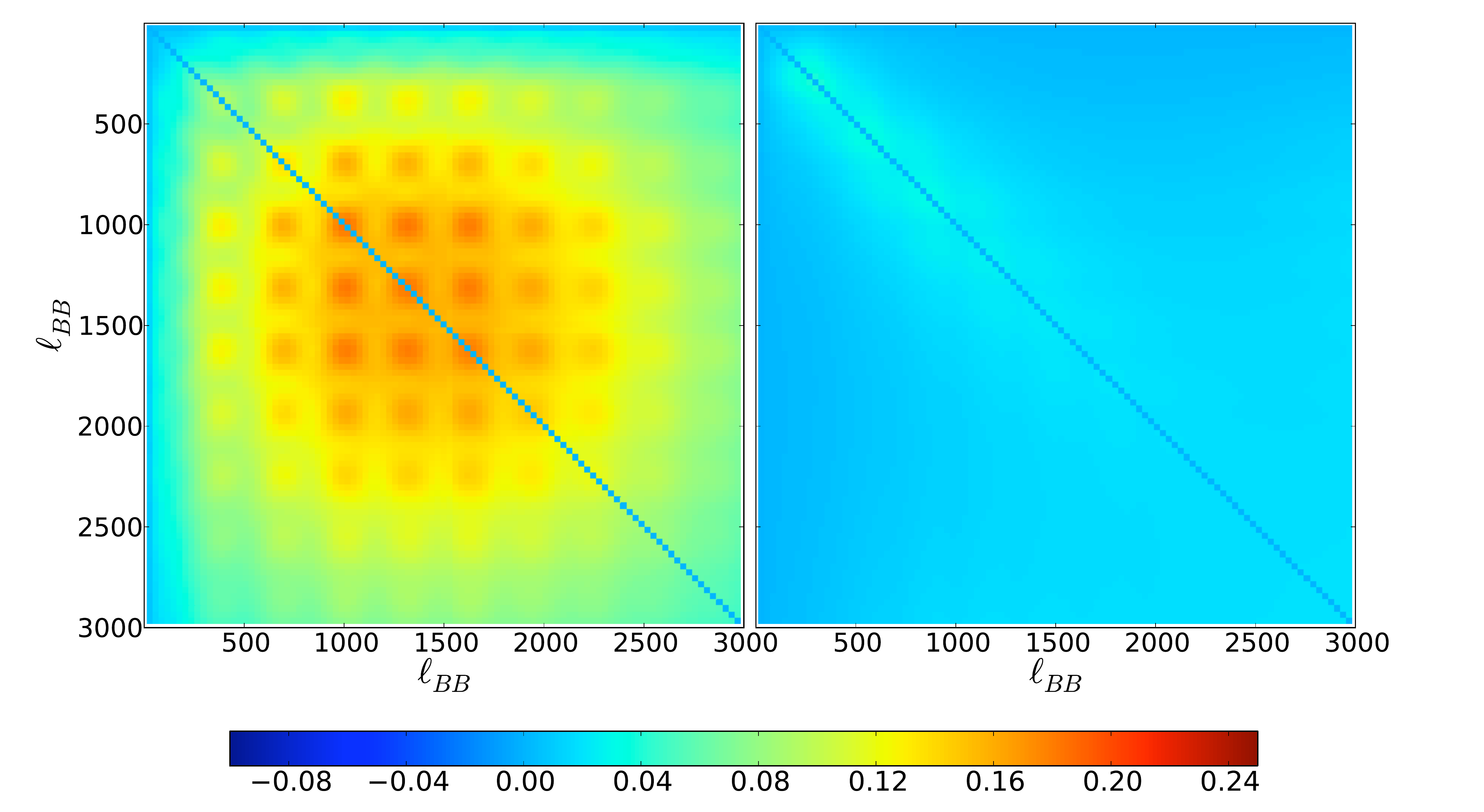}
\end{center}
\caption{\label{fig:bb_subterms}
Contributions of the individual terms in Eq.~(\ref{eqn:BBBB}) to the correlation matrix between BB bandpowers.
Left: cosmic variance of the lens power (third term in Eq.~(\ref{eqn:BBBB})); 
right: cosmic variance of the unlensed $\tilde E\tilde E$ spectrum (second term).}
\end{figure}

 Let us try to interpret the terms in  Eq.~(\ref{eqn:BBBB}).  The first term is the usual unconnected
piece of the covariance that is the only term for a Gaussian random field.   We will
loosely refer to this term as the ``Gaussian piece".    The second and third terms involve the fact that the $B$ field is constructed out of an unlensed $\tilde E$ field and the lens potential field $\phi$.   In the second term, two $BB$ band powers are connected by the covariance
of the unlensed $\tilde E$ fields they share.  In the third term, they are connected
by the shared $\phi$ fields.  Contributions to the correlation matrix for the second and third terms are shown separately in Fig.~\ref{fig:bb_subterms}.

The second term can therefore be interpreted as the covariance in $BB$ band powers generated by cosmic variance of the unlensed $\tilde E\tilde E$ power spectrum.   The
covariance it generates is positive definite in that enhanced power in $\tilde E\tilde E$ 
leads to enhanced $BB$ across the spectrum thus correlating modes
(see Fig.~\ref{fig:bb_subterms}, right panel).

The third term is the cosmic variance of the lens power.   Here the correlation 
reflects the acoustic structure of the unlensed $\tilde E\tilde E$ power spectrum.
More power in the lenses allows more power from the acoustic peaks to transfer into
$B$-modes than the acoustic troughs (see Fig.~\ref{fig:bb_subterms}, left panel). 

Finally, we note that Eq.~(\ref{eqn:BBBB}) omits a fully connected term where the $\tilde E$ and $\phi$ fields are cross connected involving
4 unique multipoles rather than three.  These contributions tend to sum incoherently 
and are subdominant in the covariance \cite{Li07}.  We omit this term in our analytic
model.

We can use these results to model the other covariance terms.   First consider
$BB,XY$ where $XY \in TT,EE,TE$.  In this case, there are no Gaussian or unconnected terms and
\begin{eqnarray}
{\rm Cov}^{BB,XY}_{\ell_1\ell_2}& =& 
 \sum_{\ell}\Bigg(
{\partial C_{\ell_1}^{BB}  \over \partial  C_{\ell}^{\tilde E\tilde E} } 
{\rm Cov}^{\tilde E\tilde E,\tilde X\tilde Y}_{\ell \ell}
{\partial C_{\ell_2}^{XY} \over \partial  C_{\ell}^{\tilde X \tilde Y} } 
\Bigg)  \nonumber\\ &&
 + \sum_{\ell}\Bigg(  {\partial C_{\ell_1}^{BB}\over \partial  C_{\ell}^{\phi\phi} } {\rm Cov}_{\ell\ell}^{\phi\phi,\phi\phi}  {\partial C_{\ell_2}^{XY} \over \partial  C_{\ell}^{\phi\phi} } \Bigg) .
 \label{eqn:BBXY}
\end{eqnarray}
In the perturbative limit for the deflection angles, this expression exactly models
all terms in the covariance.  However, again our expression has extended validity
since the derivatives are evaluated nonlinearly with CAMB.

The case of $XY=EE$ is illustrative as there is a substantial correlation.
The cosmic variance of the unlensed $\tilde E \tilde E$ power spectrum produces
contributions along the diagonal but biased to a lower $BB$ multipole $\ell_1  < \ell_2$.
This is due to the fact that most of the power in the low multipoles of $BB$ actually comes
from where the unlensed $\tilde E \tilde E$ spectrum peaks ($\ell_1 \sim 1000$).
  In previous analytic approaches,
the term that was kept was for $\ell_2=\ell$, which is linear in $C_\ell^{\phi\phi}$.

Previous approaches have dropped the term associated with the cosmic
variance of the lens power spectrum (the second term in Eq.~(\ref{eqn:BBXY})) under the justification that it is second order
in $C_\ell^{\phi\phi}$.  In fact it is the dominant contribution to the covariance
at $\ell_1,\ell_2 \gtrsim 10^3$.  This term causes a band structure in the $EE$ dependence
of the covariance.  Increasing the power in the lenses causes more power from acoustic
peaks in $\tilde E \tilde E$ to be transformed into $BB$ power while also filling in 
power in $EE$ at the acoustic troughs.    Thus peaks in $EE$ are anticorrelated with $BB$
and troughs are correlated.

Finally, there are the cases for which $XY, WZ \in TT, TE, EE$.   These cases are in principle
more complicated in that even at the perturbative level, there are many terms that
are not associated with the cosmic variance of unlensed and lens potential power spectra.
These are terms that connect the various unlensed, lensed and lens potential multipoles
in the 4 point function.     As in the case of $BB,BB$ we can again use the perturbative
approximation as a guide.    Here, there is a cancellation between the
power spectrum covariance terms and the other terms associated with the unlensed fields
for slowly varying unlensed power spectra.  These other terms reflect the fact that
at high CMB multipole moment, the unlensed fields are all lensed by the same large scale
lens realization.   For a fixed lens, neighboring bands are anticorrelated by the exchange
of power between them.   This effect does not occur for the covariances with $BB$ since
there is no unlensed $B$ field from which power can be taken.

Given this close cancellation between terms associated with the unlensed fields, we
model only the cosmic variance of the lens power spectra in these 
cases.  For $XY, WZ \in TT, TE, EE$
\begin{eqnarray}
{\rm Cov}^{XY,WZ}_{\ell_1 \ell_2} & =& 
 {1 \over 2\ell_1+1} [ C_{\ell_1}^{XW}C_{\ell_1}^{YZ} + C_{\ell_1}^{XZ}C_{\ell_1}^{YW}
] \delta_{\ell_1,\ell_2}
\nonumber\\
 &&+ \sum_{\ell}\Bigg[  {\partial C_{\ell_1}^{XY}\over \partial  C_{\ell}^{\phi\phi} } {\rm Cov}_{\ell\ell}^{\phi\phi,\phi\phi} 
{\partial C_{\ell_2}^{WZ} \over \partial  C_{\ell}^{\phi\phi} } \Bigg] .
\label{eqn:XYWZ}
\end{eqnarray}
In these cases the covariance takes a checkerboard pattern.  For $TT,TT$ or $EE,EE$ enhanced lensing power makes modes near acoustic peaks smaller and larger near troughs.   Thus peaks are correlated with peaks, troughs with troughs, and peaks are anticorrelated with
troughs.

Combining Eqs.~(\ref{eqn:BBBB}),~(\ref{eqn:BBXY}),~(\ref{eqn:XYWZ}), we have now
developed an analytic model for the lensed CMB bandpower covariance in all cases.
Comparison with the Monte Carlo covariance from~\S\ref{ssec:simulations} shows that
the difference is typically $\lsim 10$\%, leading to discrepancies in parameter
uncertainties on the order of 5\% or less.  We will explore this in more detail
in~\S\ref{sec:ngcov}.

\section{Model Independent Lensing Information}
\label{sec:lens_info}

As mentioned in the introduction, cosmological information from CMB lensing can be
obtained either from precise measurements of lensed CMB power spectra, or by applying
lens reconstruction techniques.
In this section, we will quantify the relative amount of cosmological information
which can be obtained using these two methods, in a model-independent way which
uses Fisher information matrix techniques.

In \S \ref{sec:fisher}, we review the Fisher matrix formalism as applied to the lensing potential
power spectrum.   In \S \ref{sec:KL} we construct the  Karhunen-Lo{\`e}ve (KL) basis to 
consider the relative information content. In \S \ref{CV_test}, we illustrate the necessity to take into account the non-Gaussian terms computed in \S \ref{sec:pscov}. Finally in \S \ref{Finitenoise}, we apply our formalism to realistic CMB experiments.

\subsection{Fisher Information} \label{sec:fisher}

The Fisher information matrix quantifies the information in a given data set whose
covariance matrix is known on a set of parameters $p_\alpha$ of interest. 
In order to quantify the lensing information in a model independent manner, instead of
taking cosmological parameters we take the power spectrum $C_\ell^{\phi\phi}$ itself
as the parameters of interest.  The effect of any cosmological parameter of present or
future interest can be thought of as a specific sum of these parameters.  
Rather than taking every $\ell$ as a parameter, we follow 
 \cite{Smith06} and implicitly assume that the power spectrum is smooth in $\ell$ 
 so that we can approximate it with binned perturbations around the fiducial model.
For each bin $\alpha$ in $C_\ell^{\phi\phi}$, we define a parameter $p_\alpha$ by
\be
\ln C_\ell^{\phi\phi}=\ln {C_\ell^{\phi\phi}}{|_{\rm{fid}}} + \sum_{\alpha=1}^{N_\phi}  p_\alpha B^{\phi,\ell}_{\alpha},   \qquad 0\leq \ell \leq \ell^{{\phi}}_{\rm{max}},
\label{def_band}
\ee
where $B^{\phi,\ell}_\alpha$ describes the banding and is defined as
\be
B^{\phi,\ell}_\alpha=
\left\{
\begin{array}{ccc}
1 &,& \ell_{\alpha}<\ell<\ell_{\alpha+1}\leq \ell_{\rm max}^\phi\\
0 &,& \mbox{otherwise}.
\end{array}
\right.
\ee
In practice, we choose $\phi$ bands with width $\Delta\ell=20$ 
(i.e.~$N_\phi=40$ for $\ell_{\rm max}^\phi=2000$ and $N_\phi=60$ for $\ell_{\rm max}^\phi=3000$).
We define bands in $\ln C_\ell^{\phi\phi}$ so that the power spectrum
remains positive definite for large deviations.

Note that  any cosmological parameter variation that predicts a sufficiently smooth deviation from the
fiducial model of  $\delta \ln C_\ell^{\phi\phi}$ can be represented in these parameters as
\begin{equation}
p_\alpha =\frac{1}{(\Delta \ell)_\alpha} \sum_\ell \delta \ln C_\ell^{\phi\phi} B_\alpha^{\phi,\ell}
\label{def_pa}
\end{equation}
where $(\Delta\ell)_\alpha$ is the width of bin $\alpha$.

In general, given some data vector $D_I$ which depends on parameters $p_\alpha$, the Fisher 
matrix is given by
\begin{equation}
F_{\alpha\beta} = 
\sum_{IJ}\left(\frac{\partial D_I}{\partial p_\alpha}\right)({\rm Cov}^{D,D}_{IJ})^{-1}\left(\frac{\partial D_J}{\partial p_\beta}\right).
\end{equation}
We define a Fisher matrix $F_{\alpha\beta}^P$ by specializing to the case where the parameters $p_\alpha$
are the $\phi$ bandpowers defined in Eq.~(\ref{def_pa}) and the data vector $D_I$ is the set of lensed
CMB bandpowers $D^{XY}_i$ (where $XY\in$ $TT$, $TE$, $EE$, $BB$) defined in Eq.~(\ref{eqn:CMBband}).
The derivatives $(\partial D_I/\partial p_\alpha)$ appearing in the Fisher matrix are computed
nonlinearly using CAMB, as described in the previous section.

Unless otherwise stated, our model for the covariance matrix ${\rm Cov}^{D,D}_{IJ}$ will be
based on the semi-analytical model from \S\ref{ana_approx}. (We show below that using the Monte Carlo covariance matrix from \S\ref{ssec:simulations} gives essentially identical parameter
uncertainties.)
We modify this all sky, cosmic variance limited covariance in two ways.
First we include the
possibility that the measurements contain Gaussian noise terms by replacing the
Gaussian diagonal elements with
\beqn
{\rm Cov}_{\ell\ell}^{XY,X'Y'} &=&\frac{1}{2\ell+1}\left[  (C_\ell^{XX^\prime}+ N_\ell^{XX^\prime})(C_\ell^{YY^\prime}+N_\ell^{YY^\prime}) \right.\nonumber \\ 
&& \left.+(C_\ell^{XY^\prime}+N_\ell^{XY^\prime})(C_\ell^{YX^\prime}+N_\ell^{YX^\prime})   \right],
\label{FishP}
\eeqn
where $N_\ell^{XX'}$ is the noise (cross) power spectrum.
Second, we rescale the whole resulting matrix by $1/f_{\rm sky}$ where
$f_{\rm sky}$ is the fraction of sky covered by the data set under the usual assumption
that the survey is sufficiently large that correlations across $\Delta \ell \sim 2\pi/\theta_{\rm survey}$ induced by the fundamental mode of the survey or sky cuts are irrelevant across the acoustic separation.
Also we are mainly interested in the effects of the non-Gaussian terms in the CMB covariance matrix at high $\ell$.   An accurate forecast of parameter constraints or an analysis on data would require a more detailed modeling of the shape of the survey ({\it e.g.} by introducing a minimum multipole $\ell_{\rm{min}}$).

The Fisher matrix $F^P_{\alpha\beta}$ defined in this way quantifies the lensing
information (in the form of constraints on the $\phi$ bandpowers $p_\alpha$) which
can be obtained from noisy measurements of the lensed CMB power spectrum.

We seek to compare this Fisher matrix to 
\be
F_{\alpha\beta}^{R}=\sum_{\ell\ell^\prime}\left(\frac{\partial C^{\phi\phi}_\ell}{\partial p_\alpha}\right)({\rm Cov}_{\ell\ell^\prime}^R)^{-1}\left(\frac{\partial C_{\ell^\prime}^{\phi\phi}}{\partial p_\beta}\right),
\label{FishR}
\ee
the Fisher matrix of a direct 
reconstruction  of $C_\ell^{\phi\phi}$.

We will make the approximation that the covariance matrix ${\rm Cov}_{\ell\ell'}^R$ of the reconstructed
$\phi$ bandpowers is given by the Gaussian expression
\be
{\rm Cov}_{\ell\ell^\prime}^R=\frac{2}{f_{{\rm sky}}(2\ell+1)} (C_\ell^{\phi\phi}+N_\ell^{\phi\phi})^2 \delta_{\ell\ell^\prime},
\label{eq:gaussian_covR}
\ee 
where $N_\ell^{\phi\phi}$ is the noise power spectrum of the reconstruction.
We use as a baseline two cases:
a hypothetical cosmic variance limited lens measurement where $N_\ell^{\phi\phi}=0$, and
the idealized reconstruction noise coming from the lens reconstruction from quadratic
combinations of CMB fields.  For details on how this reconstruction noise is calculated,
see \citep{Okamoto03}.

The Gaussian approximation~(\ref{eq:gaussian_covR}) makes several approximations which we state explicitly.
The power spectrum of the lens reconstruction contains an off-diagonal contribution from $\phi$
bandpowers \cite{Hu:2001fa}  (this is the ``$N_1$'' bias found by \cite{Kesden03}) which should be folded into the
covariance ${\rm Cov}^R_{\ell\ell'}$.
It is also possible that there are contributions from higher-order terms in $\phi$ \cite{Hu:2007bt}
(this is
the ``$N_2$'' bias found by \cite{Hanson:2010rp}); such contributions have been found to be
small for temperature-based lens reconstruction, but this has not been checked for polarization.
Finally, since the quadratic lens reconstruction is not a Gaussian field, its bandpower covariance
may differ from the Gaussian expression~(\ref{eq:gaussian_covR}).  For temperature-based lens
reconstruction, this issue has been studied in \cite{Hanson:2010rp} and the Gaussian expression
has been found to be a good approximation (after slightly modifying the estimator along the
lines of \cite{Dvorkin:2008tf}, see also \cite{Hu:2007bt} for alternate schemes), but the polarization case has not been studied.
A complete treatment of these issues would be very interesting but is outside the scope of
this paper; we will use the Gaussian approximation~(\ref{eq:gaussian_covR}) as a first-order
approximation to the exact Fisher matrix for lens reconstruction.

Note that the inverse of the Fisher matrix is an approximation for the covariance matrix of
$p_\alpha$
\be
{\rm Cov}_{\alpha\beta}=({\bf F}^{-1})_{\alpha\beta} 
\ee
for both the band power ($P$) and reconstruction ($R$) Fisher matrices.

\subsection{Karhunen-Lo{\`e}ve Modes}
\label{sec:KL}

While the $p_\alpha$ basis of lens power spectrum perturbations is complete, it is not
ideally suited for assessing the information content or analysis of data.     Measurements
of the many individual parameters would be highly noisy (and correlated, in the case
where the lensed CMB bandpowers $D_i^{XY}$ are being used as the observable).
In this section, we construct a more suitable basis whose eigenmodes are rank ordered
in the relative information between CMB bandpowers and lens reconstruction.
Moreover in the Fisher approximation, this
basis provides a small set of relevant parameters whose errors are uncorrelated
for both bandpower and direct measurements.  As a complete basis, it can be used
to study any cosmological parameter which affects the lensing potential.

To construct the eigenmodes, consider the  Karhunen-Lo{\`e}ve (KL) transform:
\be
{\rm Cov}^P_{\alpha \beta} v_{\beta}^{(k)}=\lambda^{(k)}{\rm Cov}^R_{\alpha \beta} v_{\beta}^{(k)}, 
\label{eqKL2}
\ee
where $ v_{\beta}^{(i)}$ and $\lambda^{(i)}$ are the KL eigenvectors and eigenvalues. 
We define the KL parameters $m_k$ as linear combinations of the band perturbations or cosmological power spectrum deviations:
\begin{eqnarray}
m_k &=&  \sum_\alpha v^{(k)}_\alpha p_\alpha \nonumber\\
&=& \sum_\alpha v^{(k)}_\alpha \frac{1}{(\Delta\ell)_\alpha} \sum_\ell \delta \ln C_\ell^{\phi\phi} B_\alpha^{\phi,\ell}.
\label{eq:mk}
\end{eqnarray}

These KL modes have the property that their covariance, either measured from the lensed power spectra or from the reconstruction are diagonal and related by the KL eigenvalues. 
The eigenvectors are normalized such that all modes have unit variance for the
direct reconstruction 
\begin{equation}
{\rm Cov}^R_{kk'}=  \delta_{kk^\prime} 
\end{equation}
and the KL construction then says that the eigenvalues are the relative variance from
the CMB bandpowers
\begin{eqnarray}
{\rm Cov}^P_{kk'}&=& \lambda ^{(k)} \delta_{kk^\prime}.
\end{eqnarray}

The KL eigenvalues are therefore the ratio of the two covariances, and give a simple quantitative way to determine which provides more information. If $\lambda^{(k)}<1$, the corresponding KL mode $m_k$ is better constrained by the power spectra than by the direct reconstruction.

\subsection{CMB and Lens Cosmic Variance\label{CV_test}}

The KL construction allows a powerful test of physical self consistency of the
bandpower covariance.  If we consider an idealized measurement in which both
the lensed CMB and the lensing potential are cosmic variance limited 
(i.e.~$N_\ell^{XY}=N_\ell^{\phi\phi}=0$) then all KL eigenvalues must be $\ge 1$,
since there cannot be more information in the lensed CMB bandpowers than 
the reconstruction. We also consider in this section that the full sky is observed, {\it i.e.} $f_{sky}=1$.

If we treat the lensed $B$-mode as a Gaussian field (i.e.~keep only the first
term in the $BB$ covariance~(\ref{eqn:BBBB})) then there are KL eigenvalues that strongly violate this physicality bound.  For example, if we suppose that the $BB$ power spectrum is measured
to $\ell_{\rm max}=2000$, then we find $\lambda_{\rm min}=0.1$.
This problem disappears when we include the full bandpower covariance: we find
$\lambda_{\rm min}=1.4$, showing that our covariance model passes this consistency test.
These results are in agreement with \cite{Hu02_synergy,Smith:2004up}.

At $\ell_{\rm max}=3000$, we find that all the polarization-related non-Gaussian covariances must be included in order
to satisfy the physicality bound $\lambda_{\rm min} \ge 1$.
For example, let us suppose that only $EE$ and $BB$ power spectra are measured (including $TT$ and $TE$
would only strengthen the example).
If we make the Gaussian approximation for Cov$^{EE,EE}$, but use non-Gaussian values for Cov$^{EE,BB}$
and Cov$^{BB,BB}$, then we find $\lambda_{\rm min}=0.9$ and fail the consistency test.
Analogously, if we make the Gaussian approximation for Cov$^{EE,BB}$, but use non-Gaussian Cov$^{EE,EE}$
and Cov$^{BB,BB}$, then we find {$\lambda_{\rm min}=0.8$} and fail.
When we include the full non-Gaussian covariance model from \S\ref{sec:pscov}, then we do not find
any violation of physicality, even when all bandpowers $TT$, $TE$, $EE$, $BB$ are included. In that case, considering all the covariance as Gaussian leads to $\lambda_{\rm min}= 0.07$.  When the full non-Gaussian covariance from \S\ref{sec:pscov} is used, we have $\lambda_{\rm min}= 1.09$ for the analytic covariance results, and {$\lambda_{\rm min}= 1.11$} when we use the covariance matrices computed from the simulations. We thus expect cosmological parameter errors to be modeled to better than a few percent for $\ell_{\rm max}\le 3000$. 
We quantify this expectation for parameter examples in \S \ref{sec:ngcov}. 
Our model captures the essential of the non-Gaussian structure of the lensed power spectra covariance.
 Most terms in this covariance model have been neglected in previous studies.

\subsection{Finite Noise}
\label{Finitenoise}
While with a perfect reconstruction of the lensing potential power spectrum we cannot expect more information from the power spectra, considering a realistic reconstruction with a finite noise could in principle lead to some modes which are better constrained by the power spectra than by reconstruction.

In Table~\ref{tab_exp}, we show some instrumental specifications that will be used throughout this paper.
For Planck, we use the lowest three HFI frequencies with measured noise levels from \cite{HFI_DPC}, with maximum multiple $\ellmax=2000$ and $\fsky=0.8$.

We  also consider a futuristic CMB polarization satellite (denoted by ``CMBpol'') with a low noise level \mbox{($\Delta_T =1\mu$K arcmin)} and a resolution similar to the SPTpol experiment \cite{McMahon09}.
Finally, we consider a ground experiment with instrumental characteristics from CMBpol assuming a 650 deg$^2$ survey and Planck sensitivity on the remainder of the Planck region (i.e.~$\fsky=0.784$). We call this combination of Planck + ground-based experiment ``\ground."

\begin{table}
\caption{\label{tab_exp}Instrumental specifications used in this paper. Sensitivities are given in $\mu$K arcmin.}
\begin{tabular}{c|c|c|c|c|c|c}
\hline
\hline
Name&Frequency& $\Delta_T$ &$ \Delta_P $& $\theta_{\rm FWHM}$ & $f_{\rm{sky}}$& $\ell_{\rm{max}}$\\\hline
 \multirow{3}{*}{Planck}  &100 GHz   &  81 & 115 & 9.5$^\prime$& 0.8 & 2000\\
                                            &143 GHz&47& 79 & 7.1$^\prime$  & 0.8 & 2000\\
                                            &217 GHz& 71 & 122& 4.7$^\prime$ &0.8  & 2000\\\hline   
Ground based &  & 1.0 & 1.41 & 1$^\prime$ &0.016  &3000\\\hline
CMBPol &  & 1.0 & 1.41 & 1$^\prime$ &0.8  &3000\\\hline \hline
\end{tabular}

\end{table}

The instrumental noise power spectrum for a single channel is \cite{Knox95}:
\begin{equation}
N_\ell^{XX}=\left(\frac{\Delta_{XX}}{T_0}\right)^2 e^{\ell(\ell+1)\theta^2_{\rm FWHM}/8\ln 2}  ,
\end{equation}
where $XX=TT, EE, BB$.  For a multi-channel experiment, the noise power spectrum is
$N_\ell = (\sum_i N_{\ell(i)}^{-1})^{-1}$, where $N_{\ell(i)}$ is the noise power spectrum
of the $i$-th channel.

Note that even though CMBpol approaches the cosmic variance limit of the CMB,
it does not reach the cosmic variance limit of lens reconstruction.   The cosmic variance of the CMB fields themselves place an irreducible noise floor on even idealized reconstruction from quadratic estimators.

The KL eigenvalues are almost always larger than one, indicating that all the KL modes are better constrained by reconstruction than with the power spectra.  The only exception is for CMBPol with a low cut-off at $\ell_{\rm{max}}=2000$. In that case  $\lambda_{\rm{min}}=0.89$, indicating that one KL mode is slightly better constrained by lensed CMB power spectra than by lens reconstruction. This number merely reflects the fact that given the low noise and beam of the CMBPol experiment, applying a cut-off at $\ell_{\rm{max}}=2000$ degrades the ability of the lensing quadratic estimator to reconstruct the lensing potential.

\begin{figure}[t]
\includegraphics[width=1.00\columnwidth]{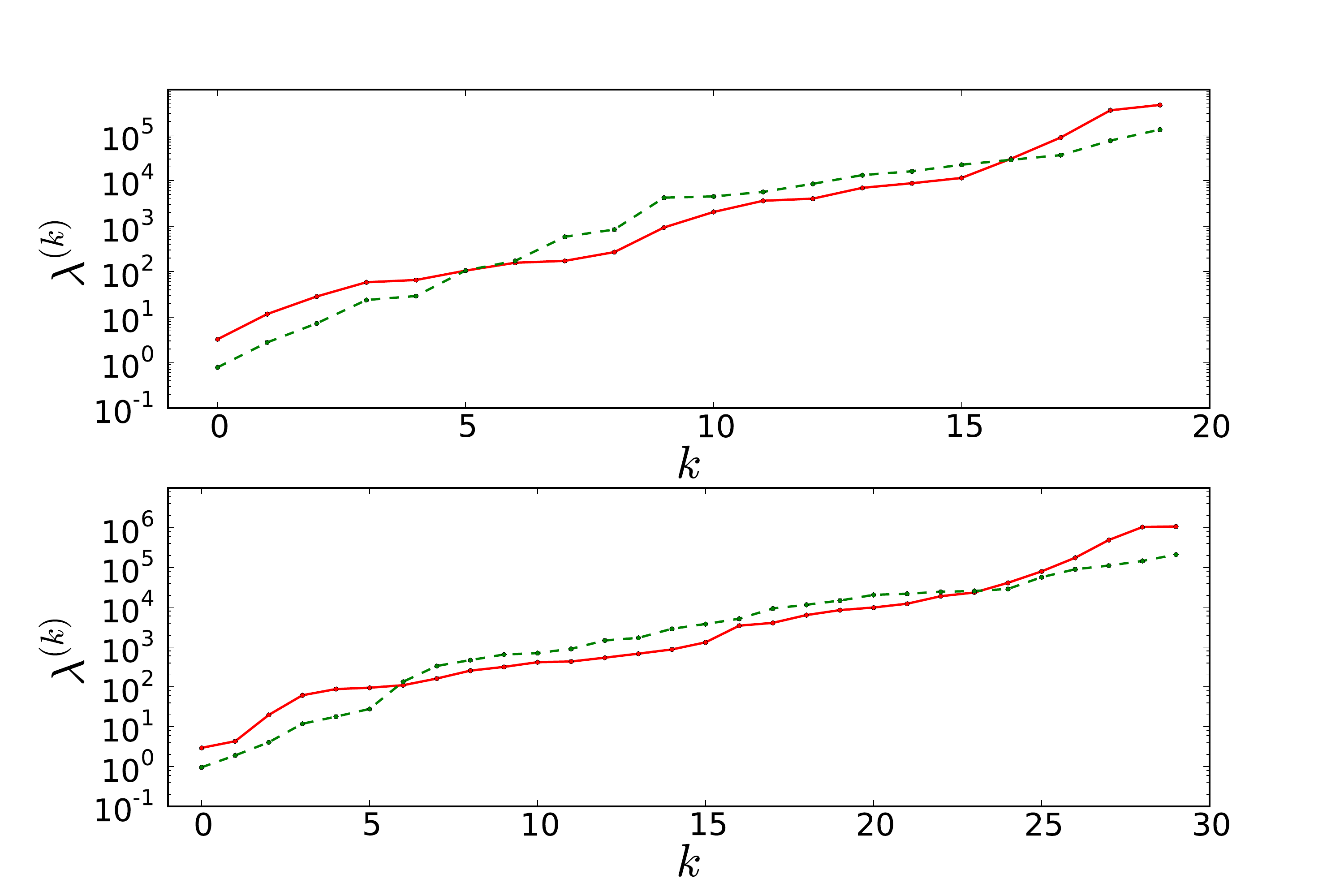}
\caption{\label{KL2_planck}KL eigenvalues for Planck (solid/red) and CMBPol (dashed/green). Top:  $\ell_{\rm{max}}=2000$,  bottom: $\ell_{\rm{max}}=3000$. }
\end{figure}

All the other KL eigenvalues are greater than one and they rapidly become much larger after the third mode (see Fig.~\ref{KL2_planck}), indicating that only one or two KL modes actually contribute to the lensed CMB power spectra, which is in agreement with Ref. \cite{Smith06}.

\section{Constraints on cosmological parameters}
\label{sec:cons}

Although the KL construction reveals extra information in lens reconstruction not available to lensed power spectra, accessing this information does not necessarily improve constraints on realistic cosmological
parameters.    Its impact depends on both how strongly and how uniquely cosmological parameter variations
change the KL mode amplitudes corresponding to the new information.

Our purpose is not to give exhaustive forecasts on cosmological parameters for various experimental configurations. {Rather, we wish to provide examples for when the extra KL information in reconstruction can and cannot make an impact.  }
 In \S~\ref{sec:valid}, we define and test a means of comparing the two in the presence of parameters that change the acoustic peaks of the unlensed CMB.  In \S~\ref{sec:comp}, we compare the errors from the power spectra to the reconstruction and we assess the impact of the non-Gaussian covariance in \S~\ref{sec:ngcov}.

\subsection{Additive Lensing Approach}
\label{sec:valid}

\begin{figure*}[t]
\includegraphics[width=1\textwidth]{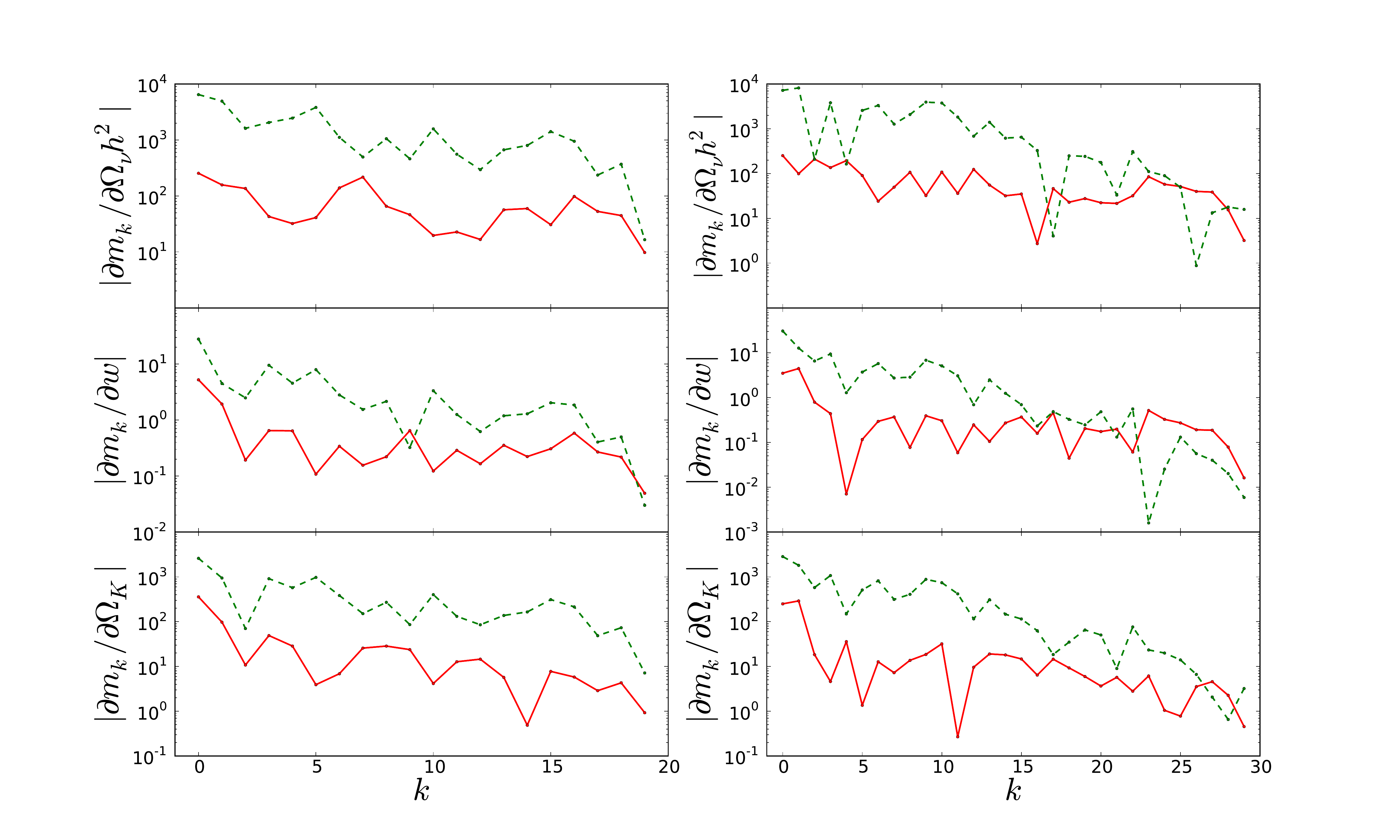}
\caption{\label{ak_planck}Derivatives of the KL modes $m_k$ with respect to $\Omega_\nu h^2$ (top), $w$ (middle), and $\Omega_K$ (bottom) with $l_{\rm{max}}=2000$ (left column) and $l_{\rm{max}}=3000$ (right column), for Planck (solid/red) and CMBpol (dashed/green).  }
\end{figure*}

The KL mode decomposition is complete and hence the errors on the mode amplitudes $m_k$ can be used to construct
the Fisher matrices $F_{cc'}^{{\rm KL},P}, F_{cc'}^{{\rm KL},R}$ corresponding to lensed power spectra and lens
reconstruction, for any set of cosmological parameters $c$:
\begin{eqnarray}
F_{cc^\prime}^{{\rm KL},P} & =&  \sum_k { 1 \over \lambda^{(k)}} {\partial m_k \over \partial c }  {\partial m_k \over \partial c^\prime },
\nonumber\\
 F_{cc^\prime}^{{\rm KL},R} & =&  \sum_k {\partial m_k \over \partial c }  {\partial m_k \over \partial c^\prime }.
 \label{Fisk_kl}
\end{eqnarray}
In other words, lensed CMB power spectrum constraints are downgraded (relative to lens reconstruction) by the
KL eigenvalues.  In practice, there are only a few eigenvalues which are not $\gg 1$, and so lensed CMB power
spectra are only sensitive to the first few eigenmodes.

The derivatives $(\partial m_k/\partial c)$ appearing above can be computed from Eq.~(\ref{eq:mk}):
\begin{equation}
{\partial m_k \over \partial c } =
\sum_\alpha v^{(k)}_\alpha \frac{1}{(\Delta\ell)_\alpha} \sum_\ell \frac{\partial \ln C_\ell^{\phi\phi}}{\partial c} B_\alpha^{\phi,\ell}.
\label{ak}
\end{equation}
Some values of these derivatives with respect to neutrino mass, dark energy equation of state, and curvature
are presented in Fig.~\ref{ak_planck}. They show a general decreasing trend but are not monotonically decreasing. For example, the 
neutrino mass derivative with $\ell_{\rm{max}}=2000$ shows a second peak at the sixth KL mode (Fig.~\ref{ak_planck}, left top panel).

The KL Fisher matrices are designed to only account for the information carried by CMB lensing and should be added to any other source of information. Indeed these matrices are highly degenerate if the parameters that control the unlensed CMB acoustic peaks  are allowed to vary. Two options are conceivable depending on the objective. If one wants to compare the ultimate amount of information on the lensing parameters carried by CMB lensing through the bandpower measurements or through a lens reconstruction, then fixing the high redshift parameters in the KL matrices is a possibility.    In this case, we expect
the KL treatment to be fully accurate within the Fisher approximation, but the resultant
error estimates are not meaningful unless other sources of information fix those parameters.

The second approach is to add other sources of information to the Fisher matrix.
The current leading source of information on these parameters is of course the acoustic
peaks themselves.   The bulk of this information comes from the unlensed CMB spectra.
To the extent that the lensing simply adds to the information in the unlensed CMB
we can approximate the total Fisher matrix as the sum 
\begin{eqnarray}
F^{{\rm KL},PU} &=&  F^{{\rm KL},P}+ F^{U} ,\nonumber\\
F^{{\rm KL},RU} &=& F^{{\rm KL},R}+ F^{U} ,   \label{eq:F_additive}
\end{eqnarray}
where $F^{U}$ is the Fisher matrix constructed out of the unlensed CMB fields with a Gaussian covariance.   We call this the ``additive lensing'' approximation.
For lensing parameters where lensing can actually destroy information in the power spectrum, such
as $\Omega_K$  \cite{Smith06} and $m_\nu$ at sufficiently large values that the neutrinos are non-relativistic at recombination, this treatment is approximate.  
On the other hand it is an approximation that affects the lensing reconstruction and power spectrum information alike.

For the power spectrum information, there is a direct check of this approximation since we
can construct the Fisher matrix from the lensed power spectra and the power spectrum
covariance
\be
F_{cc^\prime}^{{ \rm dir}, P}= \sum_{IJ}\left(\frac{\partial D_I}{\partial c}\right)({\rm{Cov}}_{IJ}^{D,D})^{-1}\left(\frac{\partial D_J}{\partial c^\prime}\right).
\label{eq:F_dir_P}
\ee
For the reconstruction, a direct check of the additive lensing approximation
would require understanding the covariance between reconstruction and power spectrum 
statistics, as well as more subtle effects such as the $N_1$ bias mentioned previously
\cite{Zahn2012}.

As an aside, note that for numerical stability
when computing the power spectrum Fisher matrices it is important to pick a parameter set $c$ where
the angular diameter distance degeneracy is manifest.  Hence in practice
derivatives with respect to cosmological parameters are computed by adjusting the Hubble parameter $h$ so that the acoustic scale is fixed when varying the values of other parameters.  Our parameter basis is then composed of three lensing parameters ($\sum m_\nu$, $w$, and $\Omega_K$) and six high-redshift parameters which control the unlensed CMB: $\{ \Omega_ch^2, \Omega_bh^2, n_s, \tau, A_s e^{-2\tau}, \theta_S \}$, where $\theta_S$ is the angle subtended by  the sound horizon at recombination.
These six high-redshift parameters are marginalized in all parameter constraints presented in this paper.

\begin{figure}[htbp]
\includegraphics[width=1.00\columnwidth]{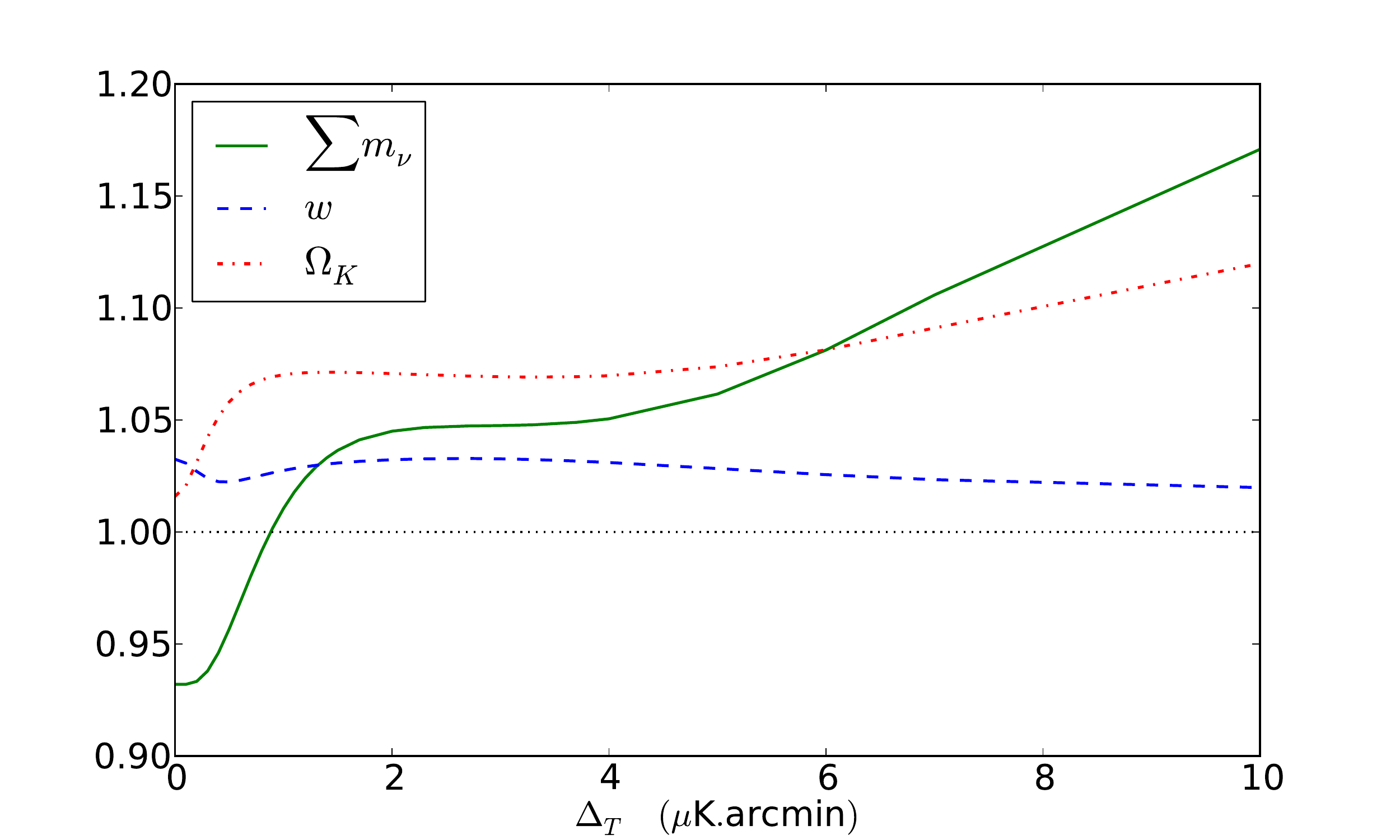}
\caption{Ratio of statistical errors computed using the additive lensing approximation (i.e.~$F^{{\rm KL},PU}$) and the exact
Fisher matrix (i.e.~$F^{{\rm dir},P}$) for individual lensing parameters, as a function
of noise level for a $\theta_{\rm{FWHM}}=1^\prime, \ell_{\rm{max}}=3000$ experiment.
The error on each lensing parameter is computed with the other two lensing parameters fixed and high-redshift parameters marginalized.}
\label{ratio2}
\end{figure}

We begin by testing the accuracy of the additive lensing approximation in the power spectrum case, where
we can simply compare the Fisher matrix $F^{{\rm KL},PU}$ obtained in the additive lensing approximation (Eq.~(\ref{eq:F_additive}))
to the exact Fisher matrix $F^{{\rm dir},P}$ (Eq.~(\ref{eq:F_dir_P})).
As can be seen in Fig.~\ref{ratio2}, which presents the ratio of the errors computed by the two different techniques for a single additional lens parameter, the two approaches are not strictly equivalent. 
As expected, this is especially true for  $\Omega_K$ where the errors from the direct lensed Fisher matrix are typically 5\% larger than those predicted by our KL formalism and can approach 20\% at high noise. For $w$, the difference in the errors is constant at about 3\% over the range of noise level considered. Finally, for $\sum m_\nu$ and the high fiducial value of 0.58eV, the agreement depends on the noise level. 
For very high levels of noise, most of the information comes from the first few peaks of the unlensed CMB and that information can be reduced by lensing.   For very low levels of noise,
the lensing information saturates to its sample variance level, while the unlensed CMB would have in principle retained information far out into the exponentially damped tail.   
In the intermediate noise regime of interest to  future CMB polarization experiments, the additive lensing approximation is accurate.  Furthermore, we have explicitly verified that as the fiducial value for $\sum m_\nu$ is lowered, the discrepancy  rapidly goes away.

\begin{figure}[htbp]
\includegraphics[width=1.00\columnwidth]{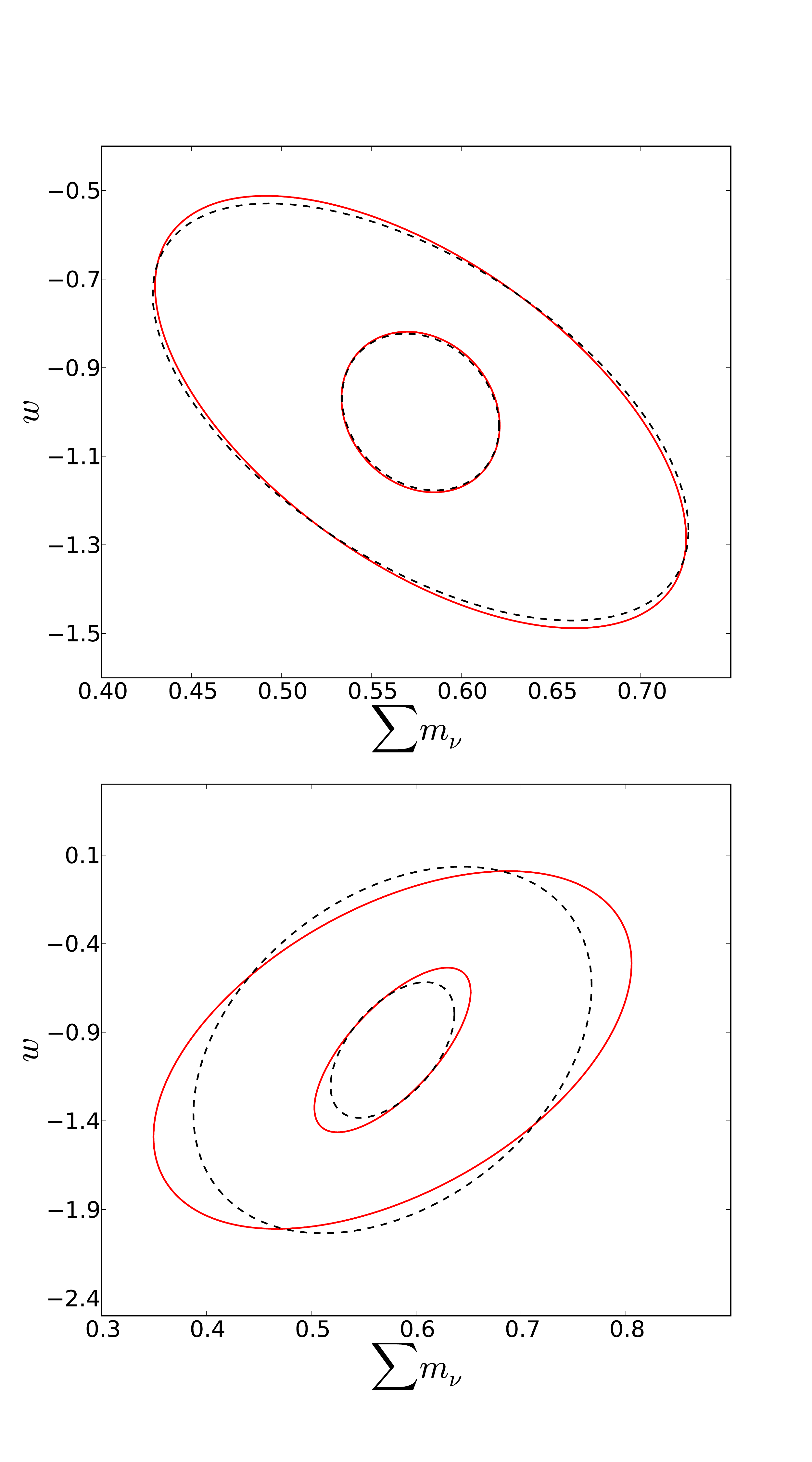}
\caption{68\% confidence limit (CL) ellipses in the ($\sum m_\nu) - w $ plane for the \ground\ (outer ellipses) and CMBpol (inner ellipses) experiments.
Errors using the additive lensing approximation ($F^{{\rm KL},PU}$) are shown in dashed/black,
and errors using the exact Fisher matrix ($F^{{\rm dir},P}$) are shown in solid/red.
Top: $\Omega_K$ is fixed. Bottom: $\Omega_K$ is marginalized.}
\label{ell2d1}
\end{figure}

The same general trends apply to cases of multiple lensing parameters.
As an example, we show in Fig.~\ref{ell2d1} (top panel), the constraints in the ($\sum m_\nu) - w $ plane ($\Omega_K$ being fixed) for the \ground\ and CMBpol experiments. The ellipses from the two approaches are in very good agreement, thus validating the additive approach in the case where curvature is fixed.

 In the case where curvature is allowed to vary (bottom panel), parameters become highly degenerate in the lensed power spectrum effects and
so the impact of unlensed information becomes larger.   For the CMBPol case
the errors in $\sigma(m_\nu)$ with $w$ and curvature marginalized are larger by a factor of 1.26  and for the \ground\ experiment they are larger by  1.20 when comparing the exact to the additive approach.   When making comparisons between reconstruction and power spectrum information in such degenerate cases with curvature, one must bear in mind these curvature induced problems \cite{Howlett2012}.

In summary, for the interesting cases where lensing provides most of the information on the
lensing parameters, the additive lensing approximation is accurate in the power spectrum
case (i.e.~the Fisher matrices $F^{{\rm KL},PU}$ and $F^{{\rm dir},P}$ agree).
The additive lensing approximation is very convenient for comparing the cosmological
information from lensed CMB power spectra and lens reconstruction.
Since the Fisher matrix is separated into a sum of unlensed and lensed contributions,
we can simply compare the Fisher matrices $F^{{\rm KL},PU}$ and $F^{{\rm KL},RU}$ defined
in Eq.~(\ref{eq:F_additive}).
This provides a metric for relative comparison of power spectrum and reconstruction
lensing information.

\subsection{Power Spectra vs. Reconstruction}
\label{sec:comp}

\begin{figure*}[t]
\includegraphics[width=1\textwidth]{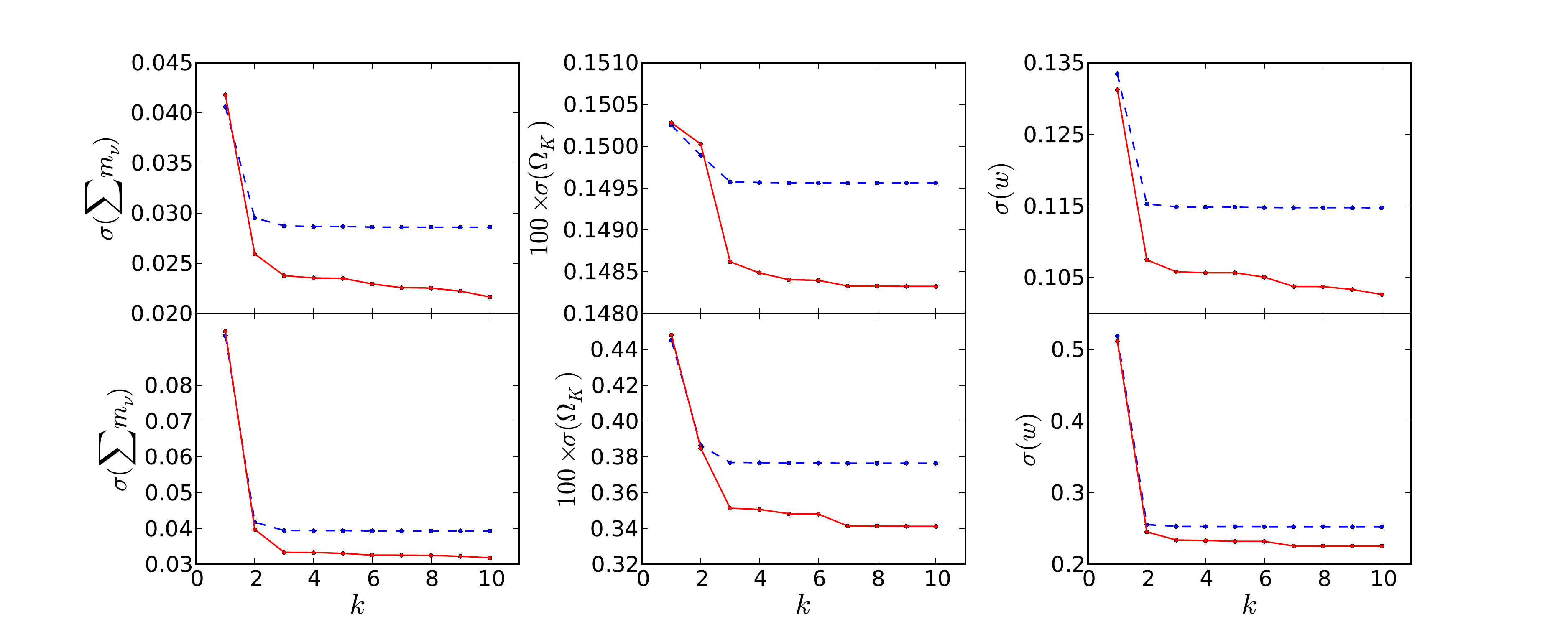}
\caption{\label{KL3} Statistical errors on $\sum m_\nu$ (left), $\Omega_K $(middle), and $w$ (right) from lens reconstruction (solid/red)
and lensed CMB power spectra (dashed/blue), using the first $k$ KL eigenmodes (where $1\le k \le 10$), for CMBpol specs with $\ellmax=3000$.
As expected, KL eigenmodes with $k \ge 3$ do not contribute to the power spectrum constraints, but can contribute
slightly in the lens reconstruction case.
Statistical errors were computed using the ``additive lensing'' approximation and the Fisher matrices
$F^{{\rm KL},PU}$ and $F^{{\rm KL},RU}$, as described in \S\ref{sec:valid}.
Top row: only one lensing parameter is varied and the other two are fixed; bottom row: all three are varied with the two not shown marginalized.
}
 \end{figure*}

\begin{figure}[htb]
\vspace{-0.4cm}
\includegraphics[width=\columnwidth]{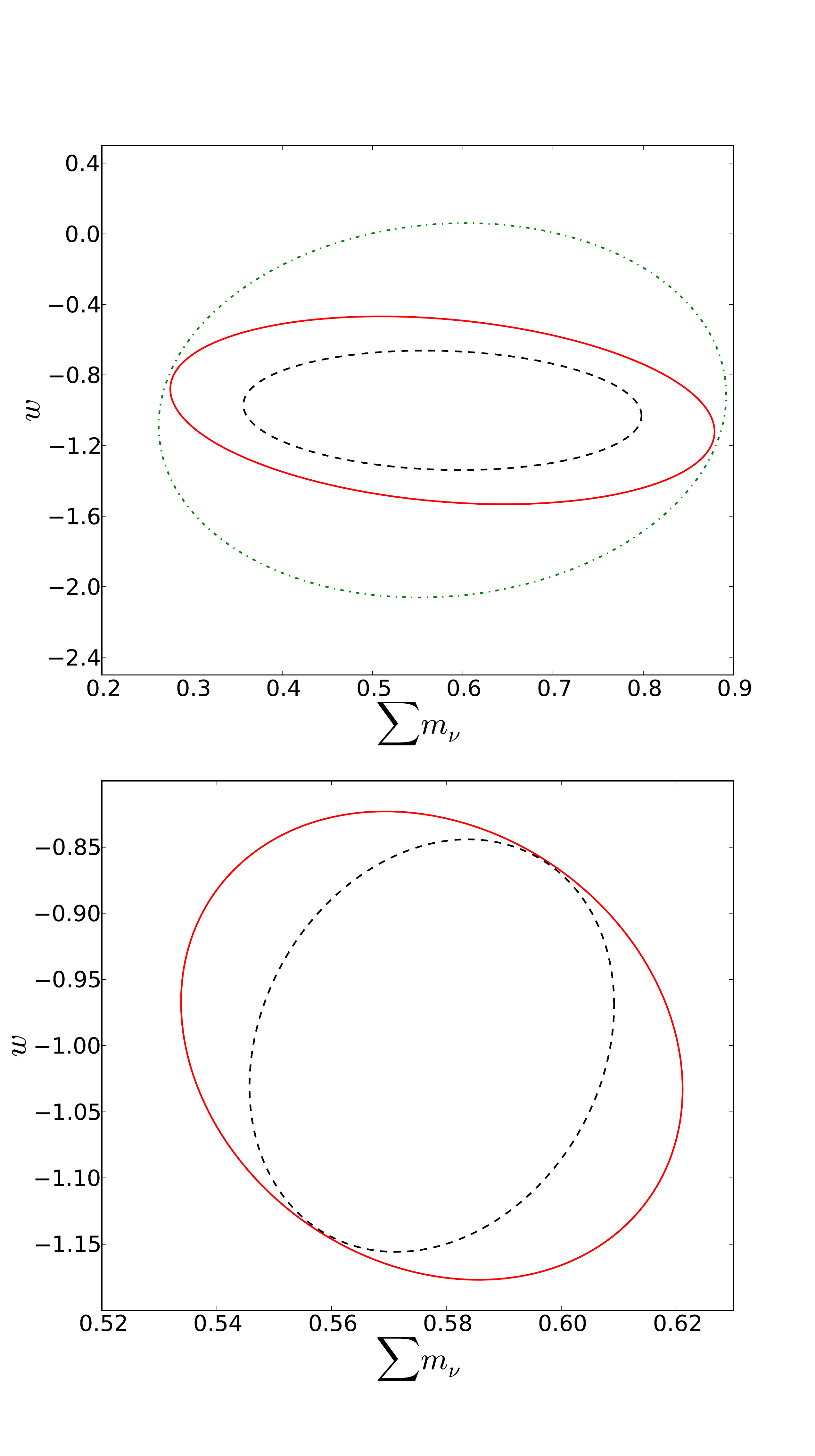}
\caption{Comparison of 68\% CL ellipses in the $(\sum m_\nu)$--$w$ plane obtained from lens reconstruction (black/dashed)
and lensed CMB power spectra (red/solid).  Statistical errors are computed using the additive lensing approximation
(i.e.~Fisher matrices $F^{{\rm KL},RU}$ and $F^{{\rm KL},PU}$) and fixed $\Omega_K$, for Planck (top panel) and CMBpol 
(bottom panel, $\ell_{\rm max}=3000$).
The outer ellipse in the top panel is {\it unlensed} CMB constraint (not shown in the bottom panel since the unlensed
constraint is much weaker than the lensed constraint).}
\label{fig:comp_PR}
\end{figure}

Given the results in the last section, we can compare cosmological parameter constraints
from lens reconstruction and lensed CMB power spectra.
Moreover, using the KL eigenmode formalism, we can explicitly verify how many KL modes actually carry the cosmological information. 

As a general statement, we find that reconstruction always carries more information than the lensed power spectra, regardless of the parameter considered. 
As the KL modes are rank-ordered in terms of highest relative information content  in the power spectra, we can choose to truncate the summation defining the KL Fisher matrices (Eq.~(\ref{Fisk_kl})) to only use the information from the first few KL modes. Those cumulative errors are presented in Fig.~\ref{KL3} (upper) for cases where there is only a single additional lensing parameter.
This case is the easiest to understand since the power spectrum information will be 
dominated by the first eigenmode.

For CMBPol, the first KL eigenvalue is close to 1 
and so the
errors for any individual lensing parameter are comparable. 
The power spectrum information saturates at 2-3 eigenmodes as expected.   These are modes for which the extra information in the reconstruction becomes manifest, but for the
three chosen lensing parameters the total impact is small for $\Omega_K$ and $w$, given the dominance of
the first mode in the derivatives in Fig.~\ref{ak_planck}.    For the neutrinos there is a somewhat
larger effect corresponding to large derivatives in both of the first two modes.  Note also that for reconstruction, the eigenmodes are not rank-ordered so some of the higher modes can
contribute more information than the lower modes.
For neutrinos, the
cumulative reduction of errors from the reconstruction (relative to the power spectrum constraint) reaches 0.7-0.8.

When more than one lensing parameter is included, resulting degeneracies can make
the information in reconstruction more important.    
However the small derivatives in  Fig.~\ref{ak_planck} still limits the practical relevance
of this information, in that other sources like the unlensed CMB and more importantly other cosmological probes quickly dominate the net information.

In Fig.~\ref{fig:comp_PR}, we show a 2-dimensional example: comparison of power spectrum and lens reconstruction constraints in the $(\sum m_\nu)$-$w$
plane with $\Omega_K$ fixed, for Planck and CMBpol.  Let us interpret this figure in light of our KL eigenmode construction.
For Planck (upper panel), there is one (roughly vertical) direction which is constrained by CMB lensing, and the lens reconstruction constraint
is stronger than the lensed power spectrum constraint.  There is another (roughly horizontal) direction which is constrained by lens reconstruction,
but very weakly constrained by {\it lensing} information in the power spectrum (it is constrained by the unlensed power spectrum).
This is consistent with the KL eigenvalues for Planck shown in Fig.~\ref{KL2_planck}: there is one KL eigenvalue which is a little larger
than 1, and the second KL eigenvalue \mbox{is $\gg 1$}.
For CMBpol (bottom panel of Fig.~\ref{fig:comp_PR}), there is one direction where the lensed power spectrum and lens reconstruction constraints are nearly
exactly equal, and another direction where lens reconstruction is somewhat better.
This is consistent with the KL eigenvalues in Fig.~\ref{KL2_planck}: the lowest KL eigenvalue is almost exactly equal to 1, and the second KL eigenvalue is
$\sim$2.

Similarly, although the KL analysis would imply that with 3 lensing parameters, there
should be substantially better lensing reconstruction constraints, for the chosen parameters
and their fiducial values the unlensed CMB information rapidly dominates.
In Fig.~\ref{KL3} (lower), we show
the impact on the parameter errors of marginalizing the other two lensing parameters. 
In this case, all three types of lensing parameters show 10\% or greater cumulative
improvements from the reconstruction due to the higher modes breaking degeneracies, but they are still of the same order of magnitude as those of the power spectra.
Further relative improvements here are limited by the unlensed CMB information which
also weakly breaks these degeneracies in the additive approach.   Note that we are somewhat
underestimating the impact of the extra reconstruction information when considering CMB-only sources of information, since this ability to break
degeneracies in the unlensed CMB is degraded by lensing.   Nonetheless the main point that
in practice the extra information accessible to lensing reconstruction with CMBPol is mainly orthogonal to realistic cosmological parameters
remains.

\subsection{Impact of non-Gaussian covariance}
\label{sec:ngcov}

\begin{figure}[t]
\includegraphics[width=\columnwidth]{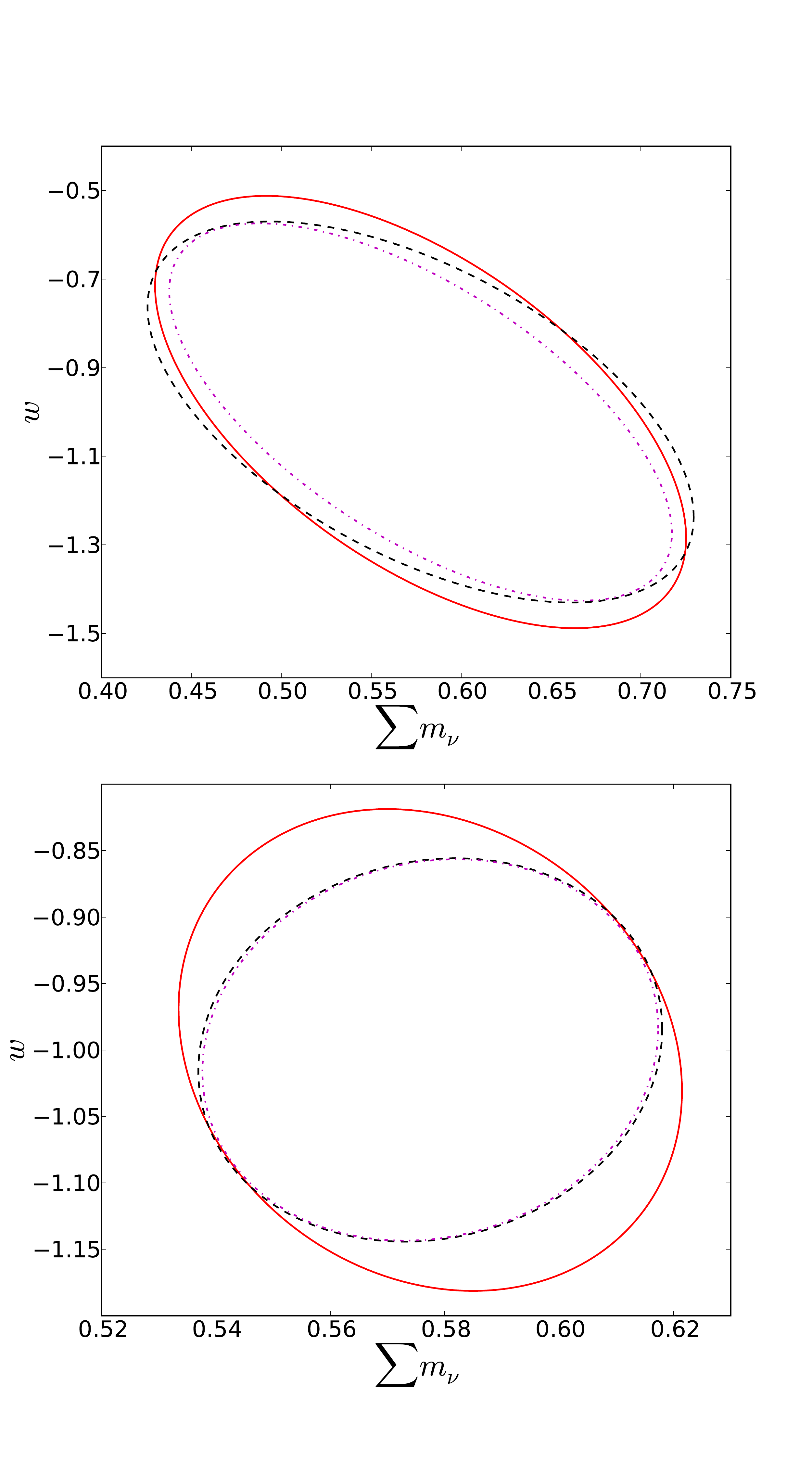}
\caption{\label{comp_NG}
68\% CL ellipses in the $(\sum m_\nu)$--$w$ plane,
computed using the exact Fisher matrix $F^{{\rm dir},P}$ with fixed $\Omega_K$,
for the \ground\ (top) and CMBPol (bottom, $\ell_{\rm max}=3000$) experiments.
Dot-dashed magenta lines are computed with the full Gaussian covariance matrix, solid red lines with the full non-Gaussian covariance from {the semi-analytical model}. Dashed black lines are computed using an intermediate covariance matrix which includes the non-Gaussian $BB$--$BB$ covariance, but Gaussian covariance for all combinations of T and E.}
\end{figure}

We conclude our analysis by investigating the impact of the non-Gaussian covariance of the lensed power spectra on the final errors on parameters. Since this does not involve reconstruction information, we work here with the direct lensed power spectra Fisher matrix $F^{\rm{dir}, P}$, rather than using the ``additive lensing'' approximation
from \S\ref{sec:valid}.

First note that if one considers the amplitude of the fiducial lensing spectrum $C_\ell^{\phi\phi}= A_{\rm lens} C_\ell^{\phi\phi}|_{\rm fid}$ as the independent lensing parameter as often done in the current literature
\cite{Sherwin11,vanEngelen12}, then non-Gaussian modeling
is required for any experiment that gains information from polarization.   This follows from our
KL treatment where the information on the amplitude comes almost exclusively from the
first mode.   
For example for CMBPol and $\ell_{\rm max}=3000$,  $\sigma(A_{\rm lens})=0.0011$ for
Gaussian covariance and $\sigma(A_{\rm lens})=0.002395$ for the non-Gaussian covariance from the simulations.   Our analytic model captures this degradation to $2.8\%$ yielding $\sigma(A_{\rm lens})=0.002329$.

If on the other hand one takes the lensing parameters as fundamental cosmological parameters, the impact of non-Gaussianity is hidden by marginalizing their impact on
$C_\ell^{\phi\phi}$.   In the previous studies where only the dominant $BB$-$BB$ covariance
was considered \cite{Smith06} the impact of non-Gaussianity on lensing parameters was
small once  the power spectra amplitude $A_s$ and the dark matter density $\Omega_c h^2$
were marginalized for $\ell_{\rm max}<2000$. 
This simplification has been employed in the subsequent literature to study parameter forecasts in a wider range of scenarios \cite{dePutter09}.

The analytic model allows us to separate out the impact of the new $EE$ and $TE$ 
covariance terms.   In Fig.~\ref{comp_NG}, we show the net impact on the $w-\sum m_\nu$
errors of including
the non-Gaussian terms on the \ground\ (top) and CMBpol (bottom) experiments, with all the high redshift parameters marginalized but $\Omega_K$ fixed.
For the \ground\ experiment the overall impact is small as one would expect from just
adding $BB$ lensing information to Planck.   For the CMBpol experiment, there is a more
substantial effect.   Interestingly this degradation is almost entirely due to the new $EE$ and
$TE$ terms and can be attributed to the use of lensing information in these spectra
to break degeneracies between lensing parameters and high redshift parameters.

These new terms can have an impact even if there is only one additional lensing
parameter.   For the CMBpol experiment if $\sum m_\nu$ and $\Omega_K$ are fixed, the error on $w$ with the full non-Gaussian covariance is $\sigma(w)=0.118$.    However, if only the non-Gaussian $BB-BB$ covariance is used, we have $\sigma(w)=0.095$, a $24$\% difference. We conclude that if in the future lensing
 information out to $\ell_{\rm max}=3000$ from the polarization fields becomes available and
 dominates parameter errors, then all of the covariance terms that involve polarization should be modeled for full accuracy.

\section{Conclusions}

We have constructed a semi-analytic model of the covariance matrix of the lensed power spectra of CMB temperature and polarization anisotropies. This model is able to reproduce the structure found in simulations of CMB lensing for the non-Gaussian terms in this covariance. More specifically, we have shown the existence and importance of second-order terms in the lensing potential that were unaccounted for in previous studies \cite{Smith06, SmithS06, Li07}.  Our model captures these effects and enables an efficient quantification of cosmological parameter errors that matches simulations to better than $\sim 3\%$ for $\ell_{\rm max}\le 3000$ even in cases where the non-Gaussianity causes an order unity degradation in the errors.

Using an parameter independent approach based on the decomposition of the information carried by CMB lensing in terms of Karhunen-Lo{\`e}ve eigenmodes, we have exhibited some cases where neglecting some of these second order terms lead to physical inconsistencies. These inconsistencies are removed once the covariance from our model is used.

We then applied the KL eigenmode technique to compare
the cosmological information that can be extracted either from measurements of the lensed power spectra or by reconstruction of the lensing potential using
quadratic estimators. 
Although the non-Gaussian covariance of the lensed spectra has no significant impact on parameter errors  for Planck, we found that it is non-negligible 
for forthcoming CMB experiments which will probe polarization at the arcminute scale.
 
If the full non-Gaussian covariance is used then there is always more information, in principle, in the reconstruction than in the lensed power spectra.  In practice, the removal of 
various biases in the reconstruction and higher order terms in the reconstruction covariance
matrix may degrade the final errors on parameters.     Furthermore, this extra information
is mainly in the detailed shape of the power spectrum of the lenses.  Typical 
cosmological parameters do not access this information as they mainly change
the amplitude of the spectrum.

The work presented here is one element in a joint and optimal likelihood analysis of CMB lensing.  Our covariance model provides a computationally efficient means of calculating
the covariance matrix of lensed CMB power spectra as a function of underlying cosmological or lens parameters.   In the future, a full joint analysis will require more accurate techniques for a similar characterization of the reconstruction covariance as well as the covariance between power spectra and reconstruction observables.

\begin{acknowledgments}
ABL wishes to thank the Kavli Institute for Cosmological Physics at the University of Chicago where this project has been initiated,  for financial support and hospitality.
KMS was supported by a Lyman Spitzer fellowship in the Department of Astrophysical Sciences at Princeton University.
Simulations in this paper were performed at the TIGRESS high performance computer center at Princeton University which is jointly 
supported by the Princeton Institute for Computational Science and Engineering and the Princeton University Office of Information Technology.  WH was supported by
Kavli Institute for Cosmological Physics at the University of Chicago through grants NSF PHY-0114422 and NSF PHY-0551142  and an endowment from the Kavli Foundation and its founder Fred Kavli,  by U.S.~Dept.\ of Energy contract DE-FG02-90ER-40560 and the David and Lucile Packard Foundation.

\end{acknowledgments}

\appendix{}

\section{Derivative approximation for non-Gaussian covariance}
\label{app:derivatives}

The covariance of $BB$ with $EE$, $TT$ or $TE$ (see Eq.~\ref{eqn:BBXY}) involves computing derivatives
of lensed CMB power spectra with respect to unlensed spectra
\begin{equation}
 \sum_{\ell}\Bigg(
{\partial C_{\ell_1}^{BB}  \over \partial  C_{\ell}^{\tilde E\tilde E} } 
{\rm Cov}^{\tilde E\tilde E,\tilde X\tilde Y}_{\ell \ell}
{\partial C_{\ell_2}^{XY} \over \partial  C_{\ell}^{\tilde X \tilde Y} } 
\Bigg) .
     \end{equation}
First note that $\partial C_{\ell_1}^{BB}/\partial C_{\ell}^{\tilde E\tilde E}$ is a
slowly varying function of both $\ell_1$ and $\ell$ given the broad kernel that transfers
power between the $\tilde E\tilde E$ and $BB$.  Thus the derivative can be well approximated by the average
response to an unlensed band perturbation of width  $\Delta L_\alpha=10$ for $\ell \in$ band
\beqn
 {\partial C_{\ell_1}^{BB}\over \partial   C_{\ell}^{\tilde E \tilde E} }   &\approx& {\partial C_{\ell_1}^{BB}\over \partial p^{\tilde E \tilde E}_\alpha }  \frac{1}{\Delta L_\alpha}  \frac{1}{  C_{\ell}^{\tilde E \tilde E}}.
\eeqn
On the other hand $\partial C_{\ell_2}^{XY} / \partial  C_{\ell}^{\tilde X \tilde Y}$
cannot in general be approximated by a band response as it will have both a
smooth piece from lensing and a $\delta_{\ell_2,\ell}$ term from the unlensed
CMB.   However, to calculate
the covariance it suffices to note that both the $BB$ derivative and the Cov term 
are slowly varying on the $\Delta L_\alpha$ scale.
Thus the sum over $\ell \in B_\alpha$ means that we can replace the true derivative with
the band average response again
\beqn
\left\langle {\partial C_{\ell_2}^{XY}\over \partial   C_{\ell}^{\tilde X \tilde Y} } \right\rangle_{\Delta L_\alpha}  &=& {\partial C_{\ell_2}^{XY}\over \partial p_\alpha^{\tilde X \tilde Y} }  \frac{1}{\Delta L_\alpha}  \frac{1}{  C_{\ell}^{\tilde X\tilde Y}}.
\eeqn
With this replacement
\begin{eqnarray}
&&
 \sum_{\ell}\Bigg(
{\partial C_{\ell_1}^{BB}  \over \partial  C_{\ell}^{\tilde E\tilde E} } 
{\rm Cov}^{\tilde E\tilde E,\tilde X\tilde Y}_{\ell \ell}
{\partial C_{\ell_2}^{XY} \over \partial  C_{\ell}^{\tilde X \tilde Y} } 
\Bigg) 
\nonumber\\
&&\approx
 \sum_\alpha \Bigg[  
\frac{\partial C_{\ell_1}^{BB}}{\partial p_\alpha^{\tilde E \tilde E}}\frac{\partial C_{\ell_2}^{XY}}{\partial p_\alpha^{\tilde X \tilde Y}} \frac{1}{(\Delta L_\alpha)^2}\sum_{\ell \in B_\alpha} \frac{ {\rm Cov}_{\ell\ell}^{\tilde E \tilde E,\tilde X \tilde Y} }{  C_{\ell}^{\tilde E \tilde E} C_{\ell}^{\tilde X \tilde Y}}
   \Bigg],
\end{eqnarray}

Note that while not necessary here, the derivatives can alternately be calculated more exactly 
by perturbing single $\ell$'s on a sparse grid in the unlensed $\ell$.   As a function of the 
unlensed $\ell$ the derivatives can be separated as
\begin{equation}
{\partial C_{\ell_1}^{XY} \over \partial  C_{\ell}^{\tilde X \tilde Y} } 
= A_{\ell} \delta_{\ell\ell_1} + B_{\ell\ell_1}
\end{equation}
into two slowly varying pieces $A_\ell$ and $B_{\ell\ell_1}$.

\bibliography{biblio}

\begin{thebibliography}{42}%
\makeatletter
\providecommand \@ifxundefined [1]{%
 \@ifx{#1\undefined}
}%
\providecommand \@ifnum [1]{%
 \ifnum #1\expandafter \@firstoftwo
 \else \expandafter \@secondoftwo
 \fi
}%
\providecommand \@ifx [1]{%
 \ifx #1\expandafter \@firstoftwo
 \else \expandafter \@secondoftwo
 \fi
}%
\providecommand \natexlab [1]{#1}%
\providecommand \enquote  [1]{``#1''}%
\providecommand \bibnamefont  [1]{#1}%
\providecommand \bibfnamefont [1]{#1}%
\providecommand \citenamefont [1]{#1}%
\providecommand \href@noop [0]{\@secondoftwo}%
\providecommand \href [0]{\begingroup \@sanitize@url \@href}%
\providecommand \@href[1]{\@@startlink{#1}\@@href}%
\providecommand \@@href[1]{\endgroup#1\@@endlink}%
\providecommand \@sanitize@url [0]{\catcode `\\12\catcode `\$12\catcode
  `\&12\catcode `\#12\catcode `\^12\catcode `\_12\catcode `\%12\relax}%
\providecommand \@@startlink[1]{}%
\providecommand \@@endlink[0]{}%
\providecommand \url  [0]{\begingroup\@sanitize@url \@url }%
\providecommand \@url [1]{\endgroup\@href {#1}{\urlprefix }}%
\providecommand \urlprefix  [0]{URL }%
\providecommand \Eprint [0]{\href }%
\providecommand \doibase [0]{http://dx.doi.org/}%
\providecommand \selectlanguage [0]{\@gobble}%
\providecommand \bibinfo  [0]{\@secondoftwo}%
\providecommand \bibfield  [0]{\@secondoftwo}%
\providecommand \translation [1]{[#1]}%
\providecommand \BibitemOpen [0]{}%
\providecommand \bibitemStop [0]{}%
\providecommand \bibitemNoStop [0]{.\EOS\space}%
\providecommand \EOS [0]{\spacefactor3000\relax}%
\providecommand \BibitemShut  [1]{\csname bibitem#1\endcsname}%
\let\auto@bib@innerbib\@empty
\bibitem [{\citenamefont {{Blanchard}}\ and\ \citenamefont
  {{Schneider}}(1987)}]{Blanchard87}%
  \BibitemOpen
  \bibfield  {author} {\bibinfo {author} {\bibfnamefont {A.}~\bibnamefont
  {{Blanchard}}}\ and\ \bibinfo {author} {\bibfnamefont {J.}~\bibnamefont
  {{Schneider}}},\ }\href@noop {} {\bibfield  {journal} {\bibinfo  {journal}
  {Astron. \& Astrophys.}\ }\textbf {\bibinfo {volume} {184}},\ \bibinfo
  {pages} {1} (\bibinfo {year} {1987})}\BibitemShut {NoStop}%
\bibitem [{\citenamefont {{Bernardeau}}(1998)}]{Bernardeau98}%
  \BibitemOpen
  \bibfield  {author} {\bibinfo {author} {\bibfnamefont {F.}~\bibnamefont
  {{Bernardeau}}},\ }\href@noop {} {\bibfield  {journal} {\bibinfo  {journal}
  {Astron. \& Astrophys.}\ }\textbf {\bibinfo {volume} {338}},\ \bibinfo
  {pages} {767} (\bibinfo {year} {1998})},\ \Eprint
  {http://arxiv.org/abs/arXiv:astro-ph/9802243} {arXiv:astro-ph/9802243}
  \BibitemShut {NoStop}%
\bibitem [{\citenamefont {{Zaldarriaga}}\ and\ \citenamefont
  {{Seljak}}(1999)}]{Zaldarriaga99}%
  \BibitemOpen
  \bibfield  {author} {\bibinfo {author} {\bibfnamefont {M.}~\bibnamefont
  {{Zaldarriaga}}}\ and\ \bibinfo {author} {\bibfnamefont {U.}~\bibnamefont
  {{Seljak}}},\ }\href {\doibase 10.1103/PhysRevD.59.123507} {\bibfield
  {journal} {\bibinfo  {journal} {Phys. Rev. D}\ }\textbf {\bibinfo {volume}
  {59}},\ \bibinfo {eid} {123507} (\bibinfo {year} {1999})},\ \Eprint
  {http://arxiv.org/abs/arXiv:astro-ph/9810257} {arXiv:astro-ph/9810257}
  \BibitemShut {NoStop}%
\bibitem [{\citenamefont {{Lewis}}\ and\ \citenamefont
  {{Challinor}}(2006)}]{Lewis06}%
  \BibitemOpen
  \bibfield  {author} {\bibinfo {author} {\bibfnamefont {A.}~\bibnamefont
  {{Lewis}}}\ and\ \bibinfo {author} {\bibfnamefont {A.}~\bibnamefont
  {{Challinor}}},\ }\href {\doibase 10.1016/j.physrep.2006.03.002} {\bibfield
  {journal} {\bibinfo  {journal} {Phys. Rep.}\ }\textbf {\bibinfo {volume}
  {429}},\ \bibinfo {pages} {1} (\bibinfo {year} {2006})},\ \Eprint
  {http://arxiv.org/abs/arXiv:astro-ph/0601594} {arXiv:astro-ph/0601594}
  \BibitemShut {NoStop}%
\bibitem [{\citenamefont {{Smith}}\ \emph {et~al.}(2007)\citenamefont
  {{Smith}}, \citenamefont {{Zahn}},\ and\ \citenamefont
  {{Dor{\'e}}}}]{Smith07}%
  \BibitemOpen
  \bibfield  {author} {\bibinfo {author} {\bibfnamefont {K.~M.}\ \bibnamefont
  {{Smith}}}, \bibinfo {author} {\bibfnamefont {O.}~\bibnamefont {{Zahn}}}, \
  and\ \bibinfo {author} {\bibfnamefont {O.}~\bibnamefont {{Dor{\'e}}}},\
  }\href {\doibase 10.1103/PhysRevD.76.043510} {\bibfield  {journal} {\bibinfo
  {journal} {Phys. Rev. D}\ }\textbf {\bibinfo {volume} {76}},\ \bibinfo
  {pages} {043510} (\bibinfo {year} {2007})},\ \Eprint
  {http://arxiv.org/abs/0705.3980} {arXiv:0705.3980} \BibitemShut {NoStop}%
\bibitem [{\citenamefont {Hirata}\ \emph {et~al.}(2008)\citenamefont {Hirata},
  \citenamefont {Ho}, \citenamefont {Padmanabhan}, \citenamefont {Seljak},\
  and\ \citenamefont {Bahcall}}]{Hirata08}%
  \BibitemOpen
  \bibfield  {author} {\bibinfo {author} {\bibfnamefont {C.~M.}\ \bibnamefont
  {Hirata}}, \bibinfo {author} {\bibfnamefont {S.}~\bibnamefont {Ho}}, \bibinfo
  {author} {\bibfnamefont {N.}~\bibnamefont {Padmanabhan}}, \bibinfo {author}
  {\bibfnamefont {U.}~\bibnamefont {Seljak}}, \ and\ \bibinfo {author}
  {\bibfnamefont {N.~A.}\ \bibnamefont {Bahcall}},\ }\href {\doibase
  10.1103/PhysRevD.78.043520} {\bibfield  {journal} {\bibinfo  {journal} {Phys.
  Rev. D}\ }\textbf {\bibinfo {volume} {78}},\ \bibinfo {pages} {043520}
  (\bibinfo {year} {2008})}\BibitemShut {NoStop}%
\bibitem [{\citenamefont {Das}\ \emph {et~al.}(2011)\citenamefont {Das},
  \citenamefont {Sherwin}, \citenamefont {Aguirre}, \citenamefont {Appel},
  \citenamefont {Bond} \emph {et~al.}}]{Das2011}%
  \BibitemOpen
  \bibfield  {author} {\bibinfo {author} {\bibfnamefont {S.}~\bibnamefont
  {Das}}, \bibinfo {author} {\bibfnamefont {B.~D.}\ \bibnamefont {Sherwin}},
  \bibinfo {author} {\bibfnamefont {P.}~\bibnamefont {Aguirre}}, \bibinfo
  {author} {\bibfnamefont {J.~W.}\ \bibnamefont {Appel}}, \bibinfo {author}
  {\bibfnamefont {J.~R.}\ \bibnamefont {Bond}},  \emph {et~al.},\ }\href
  {\doibase 10.1103/PhysRevLett.107.021301} {\bibfield  {journal} {\bibinfo
  {journal} {Phys. Rev. Lett.}\ }\textbf {\bibinfo {volume} {107}},\ \bibinfo
  {pages} {021301} (\bibinfo {year} {2011})}\BibitemShut {NoStop}%
\bibitem [{\citenamefont {Sherwin}\ \emph {et~al.}(2011)\citenamefont
  {Sherwin}, \citenamefont {Dunkley}, \citenamefont {Das}, \citenamefont
  {Appel}, \citenamefont {Bond} \emph {et~al.}}]{Sherwin11}%
  \BibitemOpen
  \bibfield  {author} {\bibinfo {author} {\bibfnamefont {B.~D.}\ \bibnamefont
  {Sherwin}}, \bibinfo {author} {\bibfnamefont {J.}~\bibnamefont {Dunkley}},
  \bibinfo {author} {\bibfnamefont {S.}~\bibnamefont {Das}}, \bibinfo {author}
  {\bibfnamefont {J.~W.}\ \bibnamefont {Appel}}, \bibinfo {author}
  {\bibfnamefont {J.}~\bibnamefont {Bond}},  \emph {et~al.},\ }\href {\doibase
  10.1103/PhysRevLett.107.021302} {\bibfield  {journal} {\bibinfo  {journal}
  {Phys.Rev.Lett.}\ }\textbf {\bibinfo {volume} {107}},\ \bibinfo {pages}
  {021302} (\bibinfo {year} {2011})},\ \Eprint {http://arxiv.org/abs/1105.0419}
  {arXiv:1105.0419 [astro-ph.CO]} \BibitemShut {NoStop}%
\bibitem [{\citenamefont {van Engelen}\ \emph {et~al.}(2012)\citenamefont {van
  Engelen}, \citenamefont {Keisler}, \citenamefont {Zahn}, \citenamefont
  {Aird}, \citenamefont {Benson} \emph {et~al.}}]{vanEngelen12}%
  \BibitemOpen
  \bibfield  {author} {\bibinfo {author} {\bibfnamefont {A.}~\bibnamefont {van
  Engelen}}, \bibinfo {author} {\bibfnamefont {R.}~\bibnamefont {Keisler}},
  \bibinfo {author} {\bibfnamefont {O.}~\bibnamefont {Zahn}}, \bibinfo {author}
  {\bibfnamefont {K.}~\bibnamefont {Aird}}, \bibinfo {author} {\bibfnamefont
  {B.}~\bibnamefont {Benson}},  \emph {et~al.},\ }\href {\doibase
  10.1088/0004-637X/756/2/142} {\bibfield  {journal} {\bibinfo  {journal}
  {Astrophys. J.}\ }\textbf {\bibinfo {volume} {756}},\ \bibinfo {eid} {142}
  (\bibinfo {year} {2012})},\ \Eprint {http://arxiv.org/abs/1202.0546}
  {arXiv:1202.0546 [astro-ph.CO]} \BibitemShut {NoStop}%
\bibitem [{\citenamefont {{Metcalf}}\ and\ \citenamefont
  {{Silk}}(1998)}]{Metcalf:1997ad}%
  \BibitemOpen
  \bibfield  {author} {\bibinfo {author} {\bibfnamefont {R.~B.}\ \bibnamefont
  {{Metcalf}}}\ and\ \bibinfo {author} {\bibfnamefont {J.}~\bibnamefont
  {{Silk}}},\ }\href {\doibase 10.1086/311080} {\bibfield  {journal} {\bibinfo
  {journal} {Astrophys. J. Lett.}\ }\textbf {\bibinfo {volume} {492}},\
  \bibinfo {pages} {L1} (\bibinfo {year} {1998})},\ \Eprint
  {http://arxiv.org/abs/arXiv:astro-ph/9710364} {arXiv:astro-ph/9710364}
  \BibitemShut {NoStop}%
\bibitem [{\citenamefont {Stompor}\ and\ \citenamefont
  {Efstathiou}(1999)}]{Stompor:1998zj}%
  \BibitemOpen
  \bibfield  {author} {\bibinfo {author} {\bibfnamefont {R.}~\bibnamefont
  {Stompor}}\ and\ \bibinfo {author} {\bibfnamefont {G.}~\bibnamefont
  {Efstathiou}},\ }\href {\doibase 10.1046/j.1365-8711.1999.02174.x} {\bibfield
   {journal} {\bibinfo  {journal} {Mon.Not.Roy.Astron.Soc.}\ }\textbf {\bibinfo
  {volume} {302}},\ \bibinfo {pages} {735} (\bibinfo {year} {1999})},\ \Eprint
  {http://arxiv.org/abs/astro-ph/9805294} {arXiv:astro-ph/9805294 [astro-ph]}
  \BibitemShut {NoStop}%
\bibitem [{\citenamefont {Hu}(2001{\natexlab{a}})}]{Hu02_synergy}%
  \BibitemOpen
  \bibfield  {author} {\bibinfo {author} {\bibfnamefont {W.}~\bibnamefont
  {Hu}},\ }\href {\doibase 10.1103/PhysRevD.65.023003} {\bibfield  {journal}
  {\bibinfo  {journal} {Phys. Rev. D}\ }\textbf {\bibinfo {volume} {65}},\
  \bibinfo {pages} {023003} (\bibinfo {year} {2001}{\natexlab{a}})}\BibitemShut
  {NoStop}%
\bibitem [{\citenamefont {Kaplinghat}\ \emph {et~al.}(2003)\citenamefont
  {Kaplinghat}, \citenamefont {Knox},\ and\ \citenamefont
  {Song}}]{Kaplinghat03}%
  \BibitemOpen
  \bibfield  {author} {\bibinfo {author} {\bibfnamefont {M.}~\bibnamefont
  {Kaplinghat}}, \bibinfo {author} {\bibfnamefont {L.}~\bibnamefont {Knox}}, \
  and\ \bibinfo {author} {\bibfnamefont {Y.-S.}\ \bibnamefont {Song}},\ }\href
  {\doibase 10.1103/PhysRevLett.91.241301} {\bibfield  {journal} {\bibinfo
  {journal} {Phys. Rev. Lett.}\ }\textbf {\bibinfo {volume} {91}},\ \bibinfo
  {pages} {241301} (\bibinfo {year} {2003})}\BibitemShut {NoStop}%
\bibitem [{\citenamefont {Acquaviva}\ and\ \citenamefont
  {Baccigalupi}(2006)}]{Acquaviva06}%
  \BibitemOpen
  \bibfield  {author} {\bibinfo {author} {\bibfnamefont {V.}~\bibnamefont
  {Acquaviva}}\ and\ \bibinfo {author} {\bibfnamefont {C.}~\bibnamefont
  {Baccigalupi}},\ }\href {\doibase 10.1103/PhysRevD.74.103510} {\bibfield
  {journal} {\bibinfo  {journal} {Phys. Rev. D}\ }\textbf {\bibinfo {volume}
  {74}},\ \bibinfo {pages} {103510} (\bibinfo {year} {2006})}\BibitemShut
  {NoStop}%
\bibitem [{\citenamefont {Hu}(2000)}]{Hu00}%
  \BibitemOpen
  \bibfield  {author} {\bibinfo {author} {\bibfnamefont {W.}~\bibnamefont
  {Hu}},\ }\href {\doibase 10.1103/PhysRevD.62.043007} {\bibfield  {journal}
  {\bibinfo  {journal} {Phys. Rev. D}\ }\textbf {\bibinfo {volume} {62}},\
  \bibinfo {pages} {043007} (\bibinfo {year} {2000})}\BibitemShut {NoStop}%
\bibitem [{\citenamefont {Benabed}\ \emph {et~al.}(2001)\citenamefont
  {Benabed}, \citenamefont {Bernardeau},\ and\ \citenamefont {van
  Waerbeke}}]{Benabed:2000jt}%
  \BibitemOpen
  \bibfield  {author} {\bibinfo {author} {\bibfnamefont {K.}~\bibnamefont
  {Benabed}}, \bibinfo {author} {\bibfnamefont {F.}~\bibnamefont {Bernardeau}},
  \ and\ \bibinfo {author} {\bibfnamefont {L.}~\bibnamefont {van Waerbeke}},\
  }\href {\doibase 10.1103/PhysRevD.63.043501} {\bibfield  {journal} {\bibinfo
  {journal} {Phys. Rev.}\ }\textbf {\bibinfo {volume} {D63}},\ \bibinfo {pages}
  {043501} (\bibinfo {year} {2001})},\ \Eprint
  {http://arxiv.org/abs/astro-ph/0003038} {arXiv:astro-ph/0003038} \BibitemShut
  {NoStop}%
\bibitem [{\citenamefont {Guzik}\ \emph {et~al.}(2000)\citenamefont {Guzik},
  \citenamefont {Seljak},\ and\ \citenamefont {Zaldarriaga}}]{Guzik:2000ju}%
  \BibitemOpen
  \bibfield  {author} {\bibinfo {author} {\bibfnamefont {J.}~\bibnamefont
  {Guzik}}, \bibinfo {author} {\bibfnamefont {U.}~\bibnamefont {Seljak}}, \
  and\ \bibinfo {author} {\bibfnamefont {M.}~\bibnamefont {Zaldarriaga}},\
  }\href {\doibase 10.1103/PhysRevD.62.043517} {\bibfield  {journal} {\bibinfo
  {journal} {Phys. Rev.}\ }\textbf {\bibinfo {volume} {D62}},\ \bibinfo {pages}
  {043517} (\bibinfo {year} {2000})},\ \Eprint
  {http://arxiv.org/abs/astro-ph/9912505} {arXiv:astro-ph/9912505} \BibitemShut
  {NoStop}%
\bibitem [{\citenamefont {Hu}(2001{\natexlab{b}})}]{Hu:2001tn}%
  \BibitemOpen
  \bibfield  {author} {\bibinfo {author} {\bibfnamefont {W.}~\bibnamefont
  {Hu}},\ }\href {\doibase 10.1086/323253} {\bibfield  {journal} {\bibinfo
  {journal} {Astrophys.J.}\ }\textbf {\bibinfo {volume} {557}},\ \bibinfo
  {pages} {L79} (\bibinfo {year} {2001}{\natexlab{b}})},\ \Eprint
  {http://arxiv.org/abs/astro-ph/0105424} {arXiv:astro-ph/0105424 [astro-ph]}
  \BibitemShut {NoStop}%
\bibitem [{\citenamefont {{Hu}}\ and\ \citenamefont {{Okamoto}}(2002)}]{Hu02}%
  \BibitemOpen
  \bibfield  {author} {\bibinfo {author} {\bibfnamefont {W.}~\bibnamefont
  {{Hu}}}\ and\ \bibinfo {author} {\bibfnamefont {T.}~\bibnamefont
  {{Okamoto}}},\ }\href {\doibase 10.1086/341110} {\bibfield  {journal}
  {\bibinfo  {journal} {\apj}\ }\textbf {\bibinfo {volume} {574}},\ \bibinfo
  {pages} {566} (\bibinfo {year} {2002})},\ \Eprint
  {http://arxiv.org/abs/arXiv:astro-ph/0111606} {arXiv:astro-ph/0111606}
  \BibitemShut {NoStop}%
\bibitem [{\citenamefont {Hirata}\ and\ \citenamefont
  {Seljak}(2003)}]{Hirata03}%
  \BibitemOpen
  \bibfield  {author} {\bibinfo {author} {\bibfnamefont {C.~M.}\ \bibnamefont
  {Hirata}}\ and\ \bibinfo {author} {\bibfnamefont {U.}~\bibnamefont
  {Seljak}},\ }\href {\doibase 10.1103/PhysRevD.68.083002} {\bibfield
  {journal} {\bibinfo  {journal} {Phys. Rev. D}\ }\textbf {\bibinfo {volume}
  {68}},\ \bibinfo {pages} {083002} (\bibinfo {year} {2003})}\BibitemShut
  {NoStop}%
\bibitem [{\citenamefont {{Smith}}\ \emph
  {et~al.}(2006{\natexlab{a}})\citenamefont {{Smith}}, \citenamefont {{Hu}},\
  and\ \citenamefont {{Kaplinghat}}}]{Smith06}%
  \BibitemOpen
  \bibfield  {author} {\bibinfo {author} {\bibfnamefont {K.~M.}\ \bibnamefont
  {{Smith}}}, \bibinfo {author} {\bibfnamefont {W.}~\bibnamefont {{Hu}}}, \
  and\ \bibinfo {author} {\bibfnamefont {M.}~\bibnamefont {{Kaplinghat}}},\
  }\href {\doibase 10.1103/PhysRevD.74.123002} {\bibfield  {journal} {\bibinfo
  {journal} {Phys. Rev. D}\ }\textbf {\bibinfo {volume} {74}},\ \bibinfo
  {pages} {123002} (\bibinfo {year} {2006}{\natexlab{a}})},\ \Eprint
  {http://arxiv.org/abs/arXiv:astro-ph/0607315} {arXiv:astro-ph/0607315}
  \BibitemShut {NoStop}%
\bibitem [{\citenamefont {{Smith}}\ \emph
  {et~al.}(2006{\natexlab{b}})\citenamefont {{Smith}}, \citenamefont
  {{Challinor}},\ and\ \citenamefont {{Rocha}}}]{SmithS06}%
  \BibitemOpen
  \bibfield  {author} {\bibinfo {author} {\bibfnamefont {S.}~\bibnamefont
  {{Smith}}}, \bibinfo {author} {\bibfnamefont {A.}~\bibnamefont
  {{Challinor}}}, \ and\ \bibinfo {author} {\bibfnamefont {G.}~\bibnamefont
  {{Rocha}}},\ }\href {\doibase 10.1103/PhysRevD.73.023517} {\bibfield
  {journal} {\bibinfo  {journal} {Phys. Rev. D}\ }\textbf {\bibinfo {volume}
  {73}},\ \bibinfo {eid} {023517} (\bibinfo {year} {2006}{\natexlab{b}})},\
  \Eprint {http://arxiv.org/abs/arXiv:astro-ph/0511703}
  {arXiv:astro-ph/0511703} \BibitemShut {NoStop}%
\bibitem [{\citenamefont {Li}\ \emph {et~al.}(2007)\citenamefont {Li},
  \citenamefont {Smith},\ and\ \citenamefont {Cooray}}]{Li07}%
  \BibitemOpen
  \bibfield  {author} {\bibinfo {author} {\bibfnamefont {C.}~\bibnamefont
  {Li}}, \bibinfo {author} {\bibfnamefont {T.~L.}\ \bibnamefont {Smith}}, \
  and\ \bibinfo {author} {\bibfnamefont {A.}~\bibnamefont {Cooray}},\ }\href
  {\doibase 10.1103/PhysRevD.75.083501} {\bibfield  {journal} {\bibinfo
  {journal} {Phys. Rev. D}\ }\textbf {\bibinfo {volume} {75}},\ \bibinfo
  {pages} {083501} (\bibinfo {year} {2007})}\BibitemShut {NoStop}%
\bibitem [{\citenamefont {{Niemack}}\ and\ \citenamefont {{\it et
  al.}}(2010)}]{ACTPol10}%
  \BibitemOpen
  \bibfield  {author} {\bibinfo {author} {\bibfnamefont {M.~D.}\ \bibnamefont
  {{Niemack}}}\ and\ \bibinfo {author} {\bibnamefont {{\it et al.}}},\ }in\
  \href {\doibase 10.1117/12.857464} {\emph {\bibinfo {booktitle} {Proc. SPIE
  Int. Soc. Opt. Eng.}}},\ \bibinfo {series} {Proc. SPIE Int. Soc. Opt. Eng.},
  Vol.\ \bibinfo {volume} {7741}\ (\bibinfo {year} {2010})\ \Eprint
  {http://arxiv.org/abs/1006.5049} {arXiv:1006.5049 [astro-ph.IM]} \BibitemShut
  {NoStop}%
\bibitem [{\citenamefont {{ B. Keating}}\ and\ \citenamefont {{\it et
  al.}}(2011)}]{Polarbear11}%
  \BibitemOpen
  \bibfield  {author} {\bibinfo {author} {\bibnamefont {{ B. Keating}}}\ and\
  \bibinfo {author} {\bibnamefont {{\it et al.}}},\ }\href@noop {} {\bibfield
  {journal} {\bibinfo  {journal} {ArXiv e-prints}\ } (\bibinfo {year}
  {2011})},\ \Eprint {http://arxiv.org/abs/1110.2101} {arXiv:1110.2101
  [astro-ph.CO]} \BibitemShut {NoStop}%
\bibitem [{\citenamefont {Lewis}\ \emph {et~al.}(2000)\citenamefont {Lewis},
  \citenamefont {Challinor},\ and\ \citenamefont {Lasenby}}]{Lewis:1999bs}%
  \BibitemOpen
  \bibfield  {author} {\bibinfo {author} {\bibfnamefont {A.}~\bibnamefont
  {Lewis}}, \bibinfo {author} {\bibfnamefont {A.}~\bibnamefont {Challinor}}, \
  and\ \bibinfo {author} {\bibfnamefont {A.}~\bibnamefont {Lasenby}},\ }\href
  {\doibase 10.1086/309179} {\bibfield  {journal} {\bibinfo  {journal}
  {Astrophys. J.}\ }\textbf {\bibinfo {volume} {538}},\ \bibinfo {pages} {473}
  (\bibinfo {year} {2000})},\ \Eprint {http://arxiv.org/abs/astro-ph/9911177}
  {arXiv:astro-ph/9911177} \BibitemShut {NoStop}%
\bibitem [{\citenamefont {{G{\'o}rski}}\ \emph {et~al.}(2005)\citenamefont
  {{G{\'o}rski}}, \citenamefont {{Hivon}}, \citenamefont {{Banday}},
  \citenamefont {{Wandelt}}, \citenamefont {{Hansen}}, \citenamefont
  {{Reinecke}},\ and\ \citenamefont {{Bartelmann}}}]{Gorski2005}%
  \BibitemOpen
  \bibfield  {author} {\bibinfo {author} {\bibfnamefont {K.~M.}\ \bibnamefont
  {{G{\'o}rski}}}, \bibinfo {author} {\bibfnamefont {E.}~\bibnamefont
  {{Hivon}}}, \bibinfo {author} {\bibfnamefont {A.~J.}\ \bibnamefont
  {{Banday}}}, \bibinfo {author} {\bibfnamefont {B.~D.}\ \bibnamefont
  {{Wandelt}}}, \bibinfo {author} {\bibfnamefont {F.~K.}\ \bibnamefont
  {{Hansen}}}, \bibinfo {author} {\bibfnamefont {M.}~\bibnamefont
  {{Reinecke}}}, \ and\ \bibinfo {author} {\bibfnamefont {M.}~\bibnamefont
  {{Bartelmann}}},\ }\href {\doibase 10.1086/427976} {\bibfield  {journal}
  {\bibinfo  {journal} {Astrophys. J.}\ }\textbf {\bibinfo {volume} {622}},\
  \bibinfo {pages} {759} (\bibinfo {year} {2005})},\ \Eprint
  {http://arxiv.org/abs/arXiv:astro-ph/0409513} {arXiv:astro-ph/0409513}
  \BibitemShut {NoStop}%
\bibitem [{\citenamefont {Lewis}(2005)}]{Lewis2005}%
  \BibitemOpen
  \bibfield  {author} {\bibinfo {author} {\bibfnamefont {A.}~\bibnamefont
  {Lewis}},\ }\href {\doibase 10.1103/PhysRevD.71.083008} {\bibfield  {journal}
  {\bibinfo  {journal} {Phys. Rev. D}\ }\textbf {\bibinfo {volume} {71}},\
  \bibinfo {pages} {083008} (\bibinfo {year} {2005})}\BibitemShut {NoStop}%
\bibitem [{\citenamefont {Smith}\ \emph {et~al.}(2004)\citenamefont {Smith},
  \citenamefont {Hu},\ and\ \citenamefont {Kaplinghat}}]{Smith:2004up}%
  \BibitemOpen
  \bibfield  {author} {\bibinfo {author} {\bibfnamefont {K.~M.}\ \bibnamefont
  {Smith}}, \bibinfo {author} {\bibfnamefont {W.}~\bibnamefont {Hu}}, \ and\
  \bibinfo {author} {\bibfnamefont {M.}~\bibnamefont {Kaplinghat}},\ }\href
  {\doibase 10.1103/PhysRevD.70.043002} {\bibfield  {journal} {\bibinfo
  {journal} {Phys. Rev.}\ }\textbf {\bibinfo {volume} {D70}},\ \bibinfo {pages}
  {043002} (\bibinfo {year} {2004})},\ \Eprint
  {http://arxiv.org/abs/astro-ph/0402442} {arXiv:astro-ph/0402442} \BibitemShut
  {NoStop}%
\bibitem [{\citenamefont {Challinor}\ and\ \citenamefont
  {Lewis}(2005)}]{Challinor:2005jy}%
  \BibitemOpen
  \bibfield  {author} {\bibinfo {author} {\bibfnamefont {A.}~\bibnamefont
  {Challinor}}\ and\ \bibinfo {author} {\bibfnamefont {A.}~\bibnamefont
  {Lewis}},\ }\href {\doibase 10.1103/PhysRevD.71.103010} {\bibfield  {journal}
  {\bibinfo  {journal} {Phys.Rev.}\ }\textbf {\bibinfo {volume} {D71}},\
  \bibinfo {pages} {103010} (\bibinfo {year} {2005})},\ \Eprint
  {http://arxiv.org/abs/astro-ph/0502425} {arXiv:astro-ph/0502425 [astro-ph]}
  \BibitemShut {NoStop}%
\bibitem [{\citenamefont {{Okamoto}}\ and\ \citenamefont
  {{Hu}}(2003)}]{Okamoto03}%
  \BibitemOpen
  \bibfield  {author} {\bibinfo {author} {\bibfnamefont {T.}~\bibnamefont
  {{Okamoto}}}\ and\ \bibinfo {author} {\bibfnamefont {W.}~\bibnamefont
  {{Hu}}},\ }\href {\doibase 10.1103/PhysRevD.67.083002} {\bibfield  {journal}
  {\bibinfo  {journal} {Phys. Rev. D}\ }\textbf {\bibinfo {volume} {67}},\
  \bibinfo {pages} {083002} (\bibinfo {year} {2003})},\ \Eprint
  {http://arxiv.org/abs/arXiv:astro-ph/0301031} {arXiv:astro-ph/0301031}
  \BibitemShut {NoStop}%
\bibitem [{\citenamefont {Hu}(2001{\natexlab{c}})}]{Hu:2001fa}%
  \BibitemOpen
  \bibfield  {author} {\bibinfo {author} {\bibfnamefont {W.}~\bibnamefont
  {Hu}},\ }\href {\doibase 10.1103/PhysRevD.64.083005} {\bibfield  {journal}
  {\bibinfo  {journal} {Phys.Rev.}\ }\textbf {\bibinfo {volume} {D64}},\
  \bibinfo {pages} {083005} (\bibinfo {year} {2001}{\natexlab{c}})},\ \Eprint
  {http://arxiv.org/abs/astro-ph/0105117} {arXiv:astro-ph/0105117 [astro-ph]}
  \BibitemShut {NoStop}%
\bibitem [{\citenamefont {{Kesden}}\ \emph {et~al.}(2003)\citenamefont
  {{Kesden}}, \citenamefont {{Cooray}},\ and\ \citenamefont
  {{Kamionkowski}}}]{Kesden03}%
  \BibitemOpen
  \bibfield  {author} {\bibinfo {author} {\bibfnamefont {M.}~\bibnamefont
  {{Kesden}}}, \bibinfo {author} {\bibfnamefont {A.}~\bibnamefont {{Cooray}}},
  \ and\ \bibinfo {author} {\bibfnamefont {M.}~\bibnamefont {{Kamionkowski}}},\
  }\href {\doibase 10.1103/PhysRevD.67.123507} {\bibfield  {journal} {\bibinfo
  {journal} {Phys. Rev. D}\ }\textbf {\bibinfo {volume} {67}},\ \bibinfo
  {pages} {123507} (\bibinfo {year} {2003})},\ \Eprint
  {http://arxiv.org/abs/arXiv:astro-ph/0302536} {arXiv:astro-ph/0302536}
  \BibitemShut {NoStop}%
\bibitem [{\citenamefont {Hu}\ \emph {et~al.}(2007)\citenamefont {Hu},
  \citenamefont {DeDeo},\ and\ \citenamefont {Vale}}]{Hu:2007bt}%
  \BibitemOpen
  \bibfield  {author} {\bibinfo {author} {\bibfnamefont {W.}~\bibnamefont
  {Hu}}, \bibinfo {author} {\bibfnamefont {S.}~\bibnamefont {DeDeo}}, \ and\
  \bibinfo {author} {\bibfnamefont {C.}~\bibnamefont {Vale}},\ }\href {\doibase
  10.1088/1367-2630/9/12/441} {\bibfield  {journal} {\bibinfo  {journal} {New
  J.Phys.}\ }\textbf {\bibinfo {volume} {9}},\ \bibinfo {pages} {441} (\bibinfo
  {year} {2007})},\ \Eprint {http://arxiv.org/abs/astro-ph/0701276}
  {arXiv:astro-ph/0701276 [astro-ph]} \BibitemShut {NoStop}%
\bibitem [{\citenamefont {Hanson}\ \emph {et~al.}(2011)\citenamefont {Hanson},
  \citenamefont {Challinor}, \citenamefont {Efstathiou},\ and\ \citenamefont
  {Bielewicz}}]{Hanson:2010rp}%
  \BibitemOpen
  \bibfield  {author} {\bibinfo {author} {\bibfnamefont {D.}~\bibnamefont
  {Hanson}}, \bibinfo {author} {\bibfnamefont {A.}~\bibnamefont {Challinor}},
  \bibinfo {author} {\bibfnamefont {G.}~\bibnamefont {Efstathiou}}, \ and\
  \bibinfo {author} {\bibfnamefont {P.}~\bibnamefont {Bielewicz}},\ }\href
  {\doibase 10.1103/PhysRevD.83.043005} {\bibfield  {journal} {\bibinfo
  {journal} {Phys.Rev.}\ }\textbf {\bibinfo {volume} {D83}},\ \bibinfo {pages}
  {043005} (\bibinfo {year} {2011})},\ \Eprint {http://arxiv.org/abs/1008.4403}
  {arXiv:1008.4403 [astro-ph.CO]} \BibitemShut {NoStop}%
\bibitem [{\citenamefont {Dvorkin}\ and\ \citenamefont
  {Smith}(2009)}]{Dvorkin:2008tf}%
  \BibitemOpen
  \bibfield  {author} {\bibinfo {author} {\bibfnamefont {C.}~\bibnamefont
  {Dvorkin}}\ and\ \bibinfo {author} {\bibfnamefont {K.~M.}\ \bibnamefont
  {Smith}},\ }\href {\doibase 10.1103/PhysRevD.79.043003} {\bibfield  {journal}
  {\bibinfo  {journal} {Phys.Rev.}\ }\textbf {\bibinfo {volume} {D79}},\
  \bibinfo {pages} {043003} (\bibinfo {year} {2009})},\ \Eprint
  {http://arxiv.org/abs/0812.1566} {arXiv:0812.1566 [astro-ph]} \BibitemShut
  {NoStop}%
\bibitem [{\citenamefont {{Planck HFI Core Team}}(2011)}]{HFI_DPC}%
  \BibitemOpen
  \bibfield  {author} {\bibinfo {author} {\bibnamefont {{Planck HFI Core
  Team}}},\ }\href {\doibase 10.1051/0004-6361/201116462} {\bibfield  {journal}
  {\bibinfo  {journal} {Astron. \& Astrophys.}\ }\textbf {\bibinfo {volume}
  {536}},\ \bibinfo {eid} {A6} (\bibinfo {year} {2011})},\ \Eprint
  {http://arxiv.org/abs/1101.2048} {arXiv:1101.2048 [astro-ph.CO]} \BibitemShut
  {NoStop}%
\bibitem [{\citenamefont {{McMahon}}\ \emph {et~al.}(2009)\citenamefont
  {{McMahon}}, \citenamefont {{Aird}}, \citenamefont {{Benson}}, \citenamefont
  {{Bleem}}, \citenamefont {{Britton}} \emph {et~al.}}]{McMahon09}%
  \BibitemOpen
  \bibfield  {author} {\bibinfo {author} {\bibfnamefont {J.~J.}\ \bibnamefont
  {{McMahon}}}, \bibinfo {author} {\bibfnamefont {K.~A.}\ \bibnamefont
  {{Aird}}}, \bibinfo {author} {\bibfnamefont {B.~A.}\ \bibnamefont
  {{Benson}}}, \bibinfo {author} {\bibfnamefont {L.~E.}\ \bibnamefont
  {{Bleem}}}, \bibinfo {author} {\bibfnamefont {J.}~\bibnamefont {{Britton}}},
  \emph {et~al.},\ }in\ \href {\doibase 10.1063/1.3292391} {\emph {\bibinfo
  {booktitle} {American Institute of Physics Conference Series}}},\ \bibinfo
  {series} {American Institute of Physics Conference Series}, Vol.\ \bibinfo
  {volume} {1185},\ \bibinfo {editor} {edited by\ \bibinfo {editor}
  {\bibnamefont {{B.~Young, B.~Cabrera, \& A.~Miller}}}}\ (\bibinfo {year}
  {2009})\ pp.\ \bibinfo {pages} {511--514}\BibitemShut {NoStop}%
\bibitem [{\citenamefont {{Knox}}(1995)}]{Knox95}%
  \BibitemOpen
  \bibfield  {author} {\bibinfo {author} {\bibfnamefont {L.}~\bibnamefont
  {{Knox}}},\ }\href {\doibase 10.1103/PhysRevD.52.4307} {\bibfield  {journal}
  {\bibinfo  {journal} {\prd}\ }\textbf {\bibinfo {volume} {52}},\ \bibinfo
  {pages} {4307} (\bibinfo {year} {1995})},\ \Eprint
  {http://arxiv.org/abs/arXiv:astro-ph/9504054} {arXiv:astro-ph/9504054}
  \BibitemShut {NoStop}%
\bibitem [{\citenamefont {Zahn}\ \emph {et~al.}()\citenamefont {Zahn},
  \citenamefont {de~Putter}, \citenamefont {Das},\ and\ \citenamefont
  {Yadav}}]{Zahn2012}%
  \BibitemOpen
  \bibfield  {author} {\bibinfo {author} {\bibfnamefont {O.}~\bibnamefont
  {Zahn}}, \bibinfo {author} {\bibfnamefont {R.}~\bibnamefont {de~Putter}},
  \bibinfo {author} {\bibfnamefont {S.}~\bibnamefont {Das}}, \ and\ \bibinfo
  {author} {\bibfnamefont {A.}~\bibnamefont {Yadav}},\ }\href@noop {} {\bibinfo
   {journal} {In prep.}\ }\BibitemShut {NoStop}%
\bibitem [{\citenamefont {{Howlett}}\ \emph {et~al.}(2012)\citenamefont
  {{Howlett}}, \citenamefont {{Lewis}}, \citenamefont {{Hall}},\ and\
  \citenamefont {{Challinor}}}]{Howlett2012}%
  \BibitemOpen
\bibfield  {journal} {  }\bibfield  {author} {\bibinfo {author} {\bibfnamefont
  {C.}~\bibnamefont {{Howlett}}}, \bibinfo {author} {\bibfnamefont
  {A.}~\bibnamefont {{Lewis}}}, \bibinfo {author} {\bibfnamefont
  {A.}~\bibnamefont {{Hall}}}, \ and\ \bibinfo {author} {\bibfnamefont
  {A.}~\bibnamefont {{Challinor}}},\ }\href {\doibase
  10.1088/1475-7516/2012/04/027} {\bibfield  {journal} {\bibinfo  {journal}
  {JCAP}\ }\textbf {\bibinfo {volume} {4}},\ \bibinfo {pages} {27} (\bibinfo
  {year} {2012})},\ \Eprint {http://arxiv.org/abs/1201.3654} {arXiv:1201.3654
  [astro-ph.CO]} \BibitemShut {NoStop}%
\bibitem [{\citenamefont {{de Putter}}\ \emph {et~al.}(2009)\citenamefont {{de
  Putter}}, \citenamefont {{Zahn}},\ and\ \citenamefont
  {{Linder}}}]{dePutter09}%
  \BibitemOpen
  \bibfield  {author} {\bibinfo {author} {\bibfnamefont {R.}~\bibnamefont {{de
  Putter}}}, \bibinfo {author} {\bibfnamefont {O.}~\bibnamefont {{Zahn}}}, \
  and\ \bibinfo {author} {\bibfnamefont {E.~V.}\ \bibnamefont {{Linder}}},\
  }\href {\doibase 10.1103/PhysRevD.79.065033} {\bibfield  {journal} {\bibinfo
  {journal} {Phys. Rev. D}\ }\textbf {\bibinfo {volume} {79}},\ \bibinfo
  {pages} {065033} (\bibinfo {year} {2009})},\ \Eprint
  {http://arxiv.org/abs/0901.0916} {arXiv:0901.0916 [astro-ph.CO]} \BibitemShut
  {NoStop}%
\end{thebibliography}%

\end{document}